# Analog regular black holes and black hole mimickers for surface-gravity waves in fluids

Valentin Pomakov [*] and Stefano Liberati [†]

*SISSA, Via Bonomea 265, 34136 Trieste, Italy,*
*INFN, Sezione di Trieste, Via Valerio 2, 34127 Trieste, Italy,*
*and Institute for Fundamental Physics of the Universe—IFPU, Via Beirut 2, 34014 Trieste, Italy*

(Received 23 April 2026; accepted 14 July 2026; published 12 August 2026)

Recent advances in the observation of black-hole candidates have renewed interest in probing their near-horizon structure and in searching for departures from the standard singular solutions of general relativity. In this context, significant effort has been devoted to regular black holes and to horizonless black-hole mimickers, motivated primarily by quantum-gravitational effects. Depending on the value of the regularization parameter relative to the object mass, typical spherically symmetric solutions can describe either of these two scenarios. Regular black-hole configurations generically feature an outer *and* an inner horizon surrounding a maximally symmetric core; the inner horizon in turn triggers mass inflation and semiclassical instabilities. The horizonless branch of the same solutions, by contrast, supports stable inner light rings when sufficiently compact, yet is itself subject to instabilities associated with long-lived quasinormal modes. Here, we investigate how to emulate these spacetimes in an analog-gravity platform based on surface-gravity waves in a shallow-water basin, with the aim of reproducing these instabilities experimentally. We begin by identifying the flow profiles and boundary conditions required to replicate the relevant effective geometries. In particular, we show that the inner-core metrics can be simulated with a nonrotating central-drainage configuration, and we propose a graded-drainage profile to connect them to an asymptotically flat exterior region. We then assess the experimental feasibility of studying the instabilities mentioned above with current technology. Our conclusion is that, while the required setup is realizable in principle, alternative media, such as Bose-Einstein condensates, may offer a more practical route to faithfully capturing the targeted physical features.

## I. INTRODUCTION

The outstanding progress in black-hole observations, through direct gravitational-wave detections and imaging with very long baseline interferometry, has renewed interest in exploring alternatives to the standard solutions of general relativity.

In particular, increasing attention has been devoted to the possibility that black holes are not singular objects but instead possess a regular interior described by a maximally symmetric geometry. Such *regular black-hole* (RBH) spacetimes are not vacuum solutions of general relativity (GR) and arise in a variety of semiclassical and quantum-gravity frameworks.

The simplest of these geometries are spherically symmetric and characterized by a single regularization scale $l$. Depending on the value of $l$, they can describe either a black hole—with a trapped region sandwiched between an outer and an inner horizon and a regular core inside the latter (often, but not necessarily, de Sitter)—or a horizonless ultracompact object featuring at least two light rings: the usual unstable outer one and a stable inner one. These are called *black hole mimickers* (BHMs). In what follows, we shall use the term *ultracompact object* (UCO) to generically refer to both of the above categories of objects.

Both of these novel structures, inner horizons and stable inner light rings, appear potentially problematic. Inner horizons, as in standard GR black holes, are generically unstable; they typically suffer from both a classical (mass-inflation) instability and semiclassical instabilities [1–7]. Inner light rings, on the other hand, correspond to minima of the effective potential and may trap and accumulate massless excitations, potentially inducing significant back-reaction on comparatively short timescales [8,9,9–12]. This expectation is not yet fully established: early numerical GR explorations [11,12] have not found clear confirmation, and proxy models (e.g., self-coupled scalar fields) have been investigated in an attempt to clarify the issue [13,14] with mixed results. What is quite firmly established is that inner light rings are associated with long-lived quasinormal modes, often viewed as a warning sign for a nonlinear instability driven by the accumulation of such slowly decaying perturbations.

[*]Contact author: vpomakov@sissa.it
[†]Contact author: liberati@sissa.it

Another interesting instability occurring in the realm of regular black holes is the blueshift instability of regular bouncing geometries studied in Ref. [15]. These spacetimes are dynamical and possess an inner and outer apparent horizon and thus a trapped region which forms and disappears in finite time, without a global event horizon or Cauchy horizon; they are globally hyperbolic, inextendible spacetimes. The inner horizon as usual induces a growing blueshift of outgoing null geodesics piling up there (the same effect propelling mass inflation and semiclassical instabilities in stationary geometries). However, in the bouncing case, this effect happens over a finite time, and hence the blueshift instability leads to a finite amplification of radiation, in contrast to the mass inflation of a Cauchy horizon, where infinities are achieved exponentially fast. Indeed, once the trapped region opens up it was shown in Ref. [15] that a large bolt of energy is released at infinity, as a consequence of this accumulation at the inner horizon. Even if they conclude that backreaction onto the background spacetime should in principle be taken into account, they suggest analog gravity as a possible setting to at least test the kinematical aspects of the problem (e.g., the bolt emission due the sudden release of the accumulated energy).

Given this picture, it would be highly valuable to study these geometries, and in particular the dynamics associated with inner horizons and inner light rings, in table-top experiments within analog-gravity platforms [16]. While analog systems can faithfully reproduce the linear field theory underlying these instabilities, they cannot directly implement the dynamical equations governing the backreaction on the geometry. Nevertheless, because the effective geometry is under experimental control, one can compare measured linear dynamics against theoretical backreaction estimates and probe the qualitative robustness of these phenomena under UV modifications. In particular, analog models typically exhibit dispersive deviations from relativistic propagation at sufficiently high frequencies, providing a natural UV cutoff. Since mass inflation is tied to arbitrarily large blueshifts, such dispersion is expected to partially tame the instability (see, e.g., Ref. [17], where the growth turns from exponential to linear for the superluminal dispersion relation). The nature of this taming will be the subject of future investigations, along with its experimental testing, which would be especially informative. In addition, the analog-gravity program could provide hints toward the end result of the instability, that is, whether the final object postinstability is a singular black hole, a regular extremal black hole, or a less compact star.

To realize this program, we must ensure that the analog-gravity laboratory setup reproduces a genuine UCO, including its concrete background spacetime structure. It is not sufficient to engineer a system that merely shares isolated features with gravitational spacetimes, such as an ergoregion or a horizon. Accordingly, this work is entirely devoted to a detailed assessment of whether, how, and under which conditions one can reproduce a background spacetime, namely, that of a static, spherically symmetric UCO with a specific mass function, using an incompressible fluid as the analog system.

The structure of the paper is as follows. After this Introduction and motivation, in Sec. II, we review relevant properties of the static spherically symmetric spacetime of interest, focusing on the role of the regularization parameter $l$ for regularizing the center, and on the core structure. In Sec. III, we review the relevant phenomenological aspects of this spacetime like instabilities of the inner horizon and of the inner light ring as well as the observables relevant to probing these. In Sec. IV, we establish the analog framework by introducing the acoustic (sonic) metric and the mapping between fluid and gravitational quantities. The main theoretical preliminaries are presented in Sec. V, before presenting a potential experimental realization of the desired UCO geometry in Sec. VI. We conclude in Sec. VII with a summary, discussion, and outlook.

## II. SPACETIME OF A STATIC SPHERICALLY SYMMETRIC ULTRACOMPACT OBJECT

Regular black-hole spacetimes are often presented as a straightforward generalization of the Schwarzschild geometry, in which the Arnowitt–Deser–Misner (ADM) mass $M$ is promoted to a radial *Misner-Sharp* mass function $m(r,l)$ characterized by a suitable regularization parameter $l$. For a static, spherically symmetric configuration, the line element can then be written as

$$ds^2 = -\left(1-\frac{2m(r,l)}{r}\right)dt^2 + \left(1-\frac{2m(r,l)}{r}\right)^{-1}dr^2 + d\Omega^2, \tag{1}$$

where as usual $d\Omega^2 = r^2(d\theta^2 + \sin^2\theta d\phi^2)$. For brevity, we will keep the dependence on $l$ implicit, writing only $m(r)$. Popular models are the Dymnikova model with $m(r) = M(1-e^{-r^3/2Ml^2})$, the Hayward model $m(r) = \frac{Mr^3}{r^3+2Ml^2}$, or the Bardeen model $m(r) = \frac{Mr^3}{(r^2+l^2)^{3/2}}$, where $M$ is the ADM mass of the spacetime. A more complete overview of models is given in Table 2 of Ref. [18].

The interpretation of $l$ as a regularization parameter becomes transparent by inspecting the Kretschmann scalar, $\mathcal{K} \equiv R_{\mu\nu\alpha\beta}R^{\mu\nu\alpha\beta}$. For a spacetime of the form (1), this is

$$\mathcal{K} = \frac{48m(r)^2}{r^6} - \frac{16m(r)}{r^5}[4m'(r) - rm''(r)] + \frac{4}{r^4}[8(m'(r))^2 - 4rm'(r)m''(r) + r^2(m''(r))^2]. \quad (2)$$

For the static, spherically symmetric geometries considered here, regularity of $\mathcal{K}$ is sufficient to ensure that the curvature remains well behaved and that the spacetime is regular [18]. From (2), it is immediate that finiteness of $\mathcal{K}$ at the origin is guaranteed provided

$$m(r \to 0) \sim r^3. \quad (3)$$

This, in turn, implies that in the $r \to 0$ limit

$$F(r,l) = 1 - \frac{2m(r,l)}{r} \simeq 1 - Cr^2, \quad (4)$$

with $C$ a constant. This is precisely the metric function of a maximally symmetric spacetime: de Sitter for $C > 0$, anti-de Sitter for $C < 0$, and Minkowski for $C = 0$. For the classes of geometries considered above, one finds $C > 0$, corresponding to a de Sitter–like core.

The dependence of $\mathcal{K}$ on $l$ for the Hayward metric is shown in Fig. 1. In particular, for Hayward, one obtains in the $r \to 0$ limit

$$\mathcal{K}(r \to 0) = \frac{24}{l^4}, \quad (5)$$

which is finite. Note that the finiteness holds for any nonvanishing $l$, although the central curvature increases rapidly for decreasing $l$, as the plot also demonstrates.

It is clear from the examples above that the metric (1) can describe a black hole with an outer and an inner horizon, an extremal configuration, or a horizonless ultracompact object, depending on the number of real positive roots of

$$F(r,l) \equiv 1 - \frac{2m(r,l)}{r}. \quad (6)$$

As an illustrative case, consider the Hayward model, $m(r) = \frac{Mr^3}{r^3+2Ml^2}$. The horizon condition $F(r,l) = 0$ reads

$$1 - \frac{2m(r)}{r} = 0 \quad \Leftrightarrow \quad r^3 - 2Mr^2 + 2Ml^2 = 0, \quad (7)$$

a cubic equation in $r$. It admits two distinct real positive roots (corresponding to the outer and inner horizons) for $0 \le l < l_*$, a degenerate (extremal) root at $l = l_*$, and no real positive roots for $l > l_*$, where

$$l_* = \frac{4M}{\sqrt{27}}. \quad (8)$$

Accordingly, the Hayward spacetime describes a regular black hole for $0 \le l \le l_*$ and a horizonless compact object for $l > l_*$ [18,19].

In the horizonless branch, one can have a black hole mimicker if its metric is characterized by a pair of light rings so that it can mimic, e.g., black hole shadows as those observed by the Event Horizon Telescope [20,21]. This is generically true for a a finite range of values of the regularization parameter, say,

$$l_* < l < l_{\rm deg}, \quad (9)$$

where the upper value $l_{\rm deg}$ corresponds to the coalescence of the two light rings into a single *degenerate* one.

In the Hayward geometry, one finds

$$l^{\rm H}_{\rm deg} \simeq 0.95M, \quad (10)$$

as can be inferred from Fig. 2 of Ref. [22]. For the Bardeen model, instead, an analytic expression is available [9],

$$l^{\rm B}_{\rm deg} = \frac{48}{25\sqrt{5}}M. \quad (11)$$

Numerically, these corresponds to

$$\text{Hayward:}\ 0.77 \lesssim l/M \lesssim 0.95, \quad (12)$$

$$\text{Bardeen:}\ 0.77 \lesssim l/M \lesssim 0.86. \quad (13)$$

for a horizonless black hole mimicker. In the remainder of this work, we will use the Hayward model as a standard example. This choice is, however, largely immaterial for our purposes, since we will mostly be interested in the limiting behavior of $m(r)$ as $r \to 0$ and $r \to \infty$, which is shared by the ultracompact-object models introduced above.

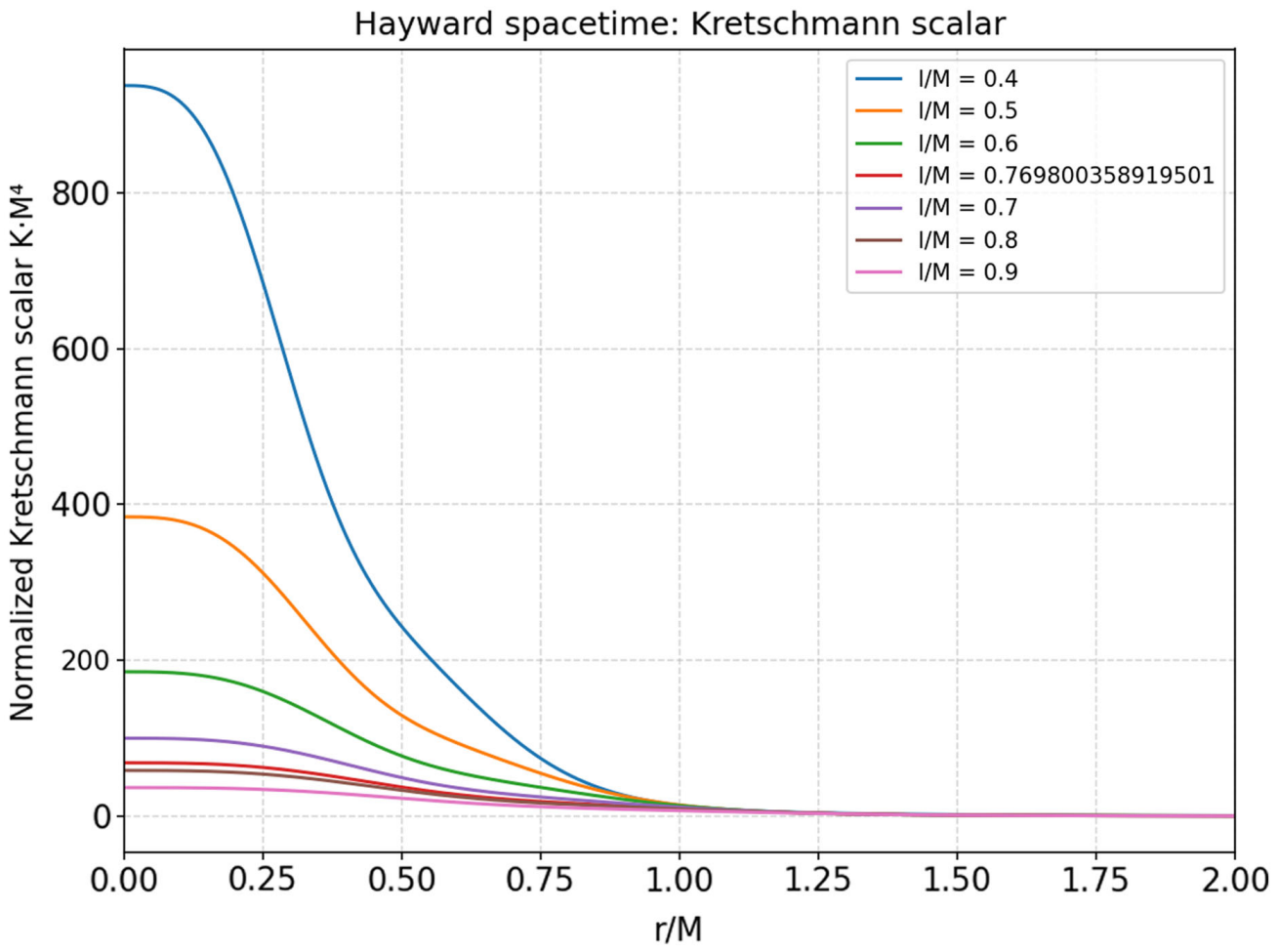

FIG. 1. The Kretschmann scalar for the Hayward spacetime in units of $M^{-4}$ as a function of radius, for different values of $l$. The middle value is $l_* = \frac{4M}{\sqrt{27}}$.

## III. RBH/BHM PHENOMENOLOGY RELEVANT FOR ANALOG GRAVITY

Let us now review a bit more in detail the kind of phenomenology that might be simulated in analog gravity experiments once the metageometry (1) is realized.

### A. Mass inflation and semiclassical instability of inner horizons

As discussed above, inner horizons of stationary black-hole geometries are well known to suffer from the *mass-inflation* instability. The key reason is that such horizons are also *future Cauchy horizons*; the entire (infinite) history of the exterior region lies in the causal past of the inner horizon, making it a Cauchy horizon for the spacetime, independently of whether a timelike singularity is present (as in the Reissner-Nordström solution) or absent. As a consequence, signals that reach the Cauchy horizon experience an unbounded blueshift. This strongly suggests that Cauchy horizons are generically unstable under classical perturbations.

The relevant perturbations can be viewed as ingoing radiation (e.g., gravitational or electromagnetic waves, or a small null flux). To translate the infinite blueshift into physical observables a standard approach consists in studying the interaction of such perturbations near the inner horizon: either two null shells (one ingoing and one outgoing) [1] or an outgoing null shell crossing a continuous ingoing null flux [23]. In both cases, the energy density to the future of the interaction grows exponentially, with an exponent controlled by the surface gravity of the inner horizon.

Accordingly, in stationary black holes with an inner/Cauchy horizon, even an arbitrarily small perturbation can generate an exponentially increasing energy density as the inner horizon is approached. The characteristic timescale is set by $1/|\kappa_-|$, where $\kappa_-$ is the inner horizon surface gravity; see, e.g., Refs. [24–26] and references therein. For standard nonextremal charged or rotating black holes, $|\kappa_-|$ is typically set by the inverse of a length scale comparable to $M$ (in geometrized units), where $M$ is the black-hole mass; thus one typically has $|\kappa_-| = \mathcal{O}(1/M)$ up to dimensionless factors depending on $Q/M$ or $a/M$. The instability therefore develops on a timescale comparable to the light-crossing time, i.e., extremely rapidly for astrophysical black holes.

While at first sight mass inflation might seem a pathology of idealized stationary solutions, related instabilities have recently been found to persist even for *dynamical* black holes with slowly evolving inner horizons. In that case, one finds an exponential accumulation of energy over a *finite* time interval. Although the growth is in this case finite, it can still be efficient enough for backreaction to become non-negligible on short timescales [15,27]. This is particularly significant because it suggest that some instability would be present also in simulated analog geometries as long as they are slowly evolving.

Finally, an additional layer of subtlety arises from a *semiclassical* instability that has been reported at the level of the renormalized stress-energy tensor. In particular, for Cauchy horizons in the Unruh vacuum, the expectation value $\langle T_{\mu\nu}\rangle_{\rm ren}$ can diverge whenever the (squared) surface gravities of the outer (event) and inner (Cauchy) horizons do not coincide, i.e., when $\kappa_-^2 \neq \kappa_+^2$; see, e.g., Refs. [5,28].

Whether dynamical inner horizons exhibit the same mechanism in full generality remains unclear. Nevertheless, on physical grounds, one may expect that, for slowly evolving horizons, an instability of this kind would manifest as an exponential growth over a *finite* interval of time like for the classical instability and could therefore impact also the semiclassical evolution of dynamical black holes. We do not expect this quantum instability to be relevant in a classical system like water-based analogs, but it could definitely play a relevant role in quantum systems like superfluid helium or Bose-Einstein condensates.

### B. BHM light ring instability

A spherical compact object admits circular null orbits (light rings) whenever there exist solutions to

$$\frac{d}{dr}\left(\frac{-g_{tt}}{r^2}\right) = 0. \tag{14}$$

It has been proven that for horizonless, regular, smooth, and asymptotically flat spacetimes this equation admits (at least) two solutions for sufficiently compact configurations. Remarkably, one of these solutions is necessarily *geometrically stable*; i.e., it corresponds to a local minimum of the effective potential governing null propagation. This was proved first assuming GR together with the null energy condition [11] and later without imposing any specific gravitational field equations [29]. This conclusion also extends to axisymmetric configurations (see, e.g., Ref. [12]). For the above-mentioned black hole mimicker geometries, this inner light ring typically lies inside the nonvacuum region of the UCO. The other solution is the familiar *geometrically unstable* outer light ring, corresponding to a local maximum of the potential. For spherically symmetric, nonrotating, UCO geometries close to a black-hole spacetime, this is located near $r \simeq 3M$ [8], i.e., the value it normally has in the Schwarzschild geometry.

Geometric stability, however, suggests a potential *dynamical* instability [8,11,12,29,30]. Because the inner light ring is a potential minimum, perturbations generate null trajectories that remain confined to radii close to the circular orbit. As a consequence, a sustained flux of perturbations can lead to an accumulation of energy in the vicinity of the inner light ring, potentially invalidating the test-field approximation and requiring one to account for backreaction on the spacetime, i.e., a genuine dynamical instability.[1]

More generally, it has been shown that in the eikonal limit long-lived perturbation modes are controlled by the light-ring structure and peak near the inner light ring. If dissipation mechanisms are neglected, this behavior of linear perturbations may trigger a fully nonlinear instability for spherically symmetric configurations [8].

It remains possible that, in some models, the relevant instability timescale exceeds the age of the Universe. However, for specific systems such as boson stars, numerical studies indicate that ultracompact configurations can indeed be unstable on astrophysically relevant timescales, evolving either toward black-hole formation or toward less compact objects [8]. Generalizing, it may be then that, in the absence of efficient dissipation mechanisms capable of quenching the energy buildup, horizonless UCOs may be generically unstable and may at best represent transient stages along the evolution toward more or less compact configurations [11,12].

### C. Echoes from BHM

Both black holes (regular or singular) and horizonless BHMs typically possess an *outer* light ring. Therefore, the mere observation of a light ring, e.g., through its shadow, does not discriminate between them; this is the essence of the "black-hole mimicker" paradigm. A more direct discriminator is the presence of *echoes* in the postmerger (or postcollapse) ringdown signal [30] or fine details in the shawdow itself (see, e.g., Refs. [31,32]).

Echoes are the time-domain counterpart of long-lived modes that are partially trapped between the potential barrier at the unstable outer light ring and an inner, reflective structure (e.g., a central centrifugal barrier), which in BHMs is associated with a potential well around the stable inner light ring [30]. Because these modes propagate deep into the interior before being reflected, echoes provide a direct probe of the core.

The echo time delay is related to the round-trip travel time of null perturbations between the outer barrier and the inner reflective region. In practice, however, for very compact objects, the dominant contribution typically comes not from the flat-space equivalent light-crossing time but from the relativistic time delay caused by large core curvatures as in Fig. 1. Consequently, the delay is closely tied to compactness (see Eq. (19) of Ref. [30]), and therefore in our case to the regularization scale.[2] In an analog-gravity setup where $l$ can be tuned, one would effectively tune core properties and, correspondingly, the echo delay, making it natural to seek an experimental analog of echoes.

However, the echo time delay alone is not a robust diagnostic. First, it is not necessarily monotonic in compactness (see Fig. 3 of Ref. [30]), so distinct compactness values can yield similar delays. Second, the core potential well can be in some models very deep [30] potentially leading to very large (and possibly unobservable) delays [30]. Third, echo modeling typically relies on a test-field approximation, whereas the very phenomenon of interest here involves backreaction as echoes are tightly connected to the presence of the inner light ring; if the object rapidly evolves (e.g., toward BH formation), the echo signal may be largely deformed [33].

For these reasons, it is essential to complement echoes, when present, with the full spectrum of quasinormal modes (QNMs), which is less affected by such degeneracies. In the horizonful (RBH) branch, echoes are not expected, and QNMs do not directly probe the inner-horizon instability since it lies behind the outer horizon. Nevertheless, the QNM spectrum *after* the instability could carry information about the nature of any new quasiequilibrium configuration.

We emphasize that the qualitative features described above depend mainly on the existence and arrangement of effective potential barriers and wells, rather than on the specific gravitational field equations. This "dynamics-independent" phenomenology is precisely what makes analog-gravity tests valuable even though the laboratory system obeys different microscopic dynamics. In particular, we expect the onset of backreaction to manifest observationally as a "sloshing" in the analog echo/QNM signal, reminiscent of what has been observed in vortex analog systems realizing black-hole-bomb-like instabilities [34].

A detailed computation of the echo/QNM signal and its experimental detection are beyond the scope of this work. Here, as a first step, we provide a detailed analysis of whether and to what extent the background spacetime (1) can be realized using surface-gravity waves in an incompressible fluid in a shallow-water basin. Analogous considerations for Bose-Einstein condensates will be presented in future work.

## IV. ACOUSTIC GEOMETRIES FOR RBH/BHM SPACETIMES

In most analog-gravity studies, experimental constraints make it impractical to reproduce a specific background spacetime in detail. Instead, one typically engineers and identifies an analog ergoregion or horizon: the minimal ingredients needed to observe phenomena such as superradiance or Hawking radiation [34–36]. The present work aims to go a step further, by assessing whether one can realize an analog of an *entire* RBH/BHM spacetime, or at least the regions relevant for the effects of interest: the inner light ring and echoes in the BHM branch and the inner-horizon instability in the RBH branch.

[1]This accumulation requires a continuous input of perturbations: there is no reason to expect spontaneous amplification of either linear or nonlinear perturbations in the absence of an ergoregion supplying negative-energy modes. Indeed, it is unclear how such an amplification without continuous input would be compatible with energy conservation when no ergoregion is present [30].

[2]The compactness $C_R$ of an object with radius $R$ is related to the regularization scale $l$ in the following way: $C_R(l) \equiv 1 - g_{rr}^{-1}(r = R, l) = \frac{2m(R,l)}{R} = \frac{2MR^2}{R^3+2Ml^2}$, where we have again used the Hayward model as an example in the last step. Reducing/increasing $l$, the compactness (and also the Kretschmann scalar at the center $r = 0$) of the spherically symmetric object increases/decreases.

Notably, both the inner horizon (in the RBH case) and the stable inner light ring (in the BHM limit) are typically located near the boundary of an approximately maximally symmetric core.[3] For this reason, our primary goal will ultimately be the realization of this core and near-core region. Nevertheless, we will also discuss the requirements for reproducing the asymptotic region $r \to \infty$. Before specializing to these two regimes, we will keep the analysis as general as possible and work with the full spacetime whenever feasible.

To realize a gravitational spacetime *unambiguously* in an analog-gravity setting, one needs a precise correspondence between the acoustic (sonic) metric and the target gravitational metric (up to conformal factors, which are irrelevant for null rays). Establishing this metric-level identification immediately provides a mapping between fluid variables and gravitational quantities.

Because the gravitational quantities of interest are functions of the areal radius $r$, some care is required in translating them into their fluid counterparts. In particular, the identification typically prescribes a specific $r$-dependence for the fluid variables, which may be experimentally unattainable because the fluid must simultaneously satisfy its own equations of motion. A concrete illustration is provided in Ref. [16], where the continuity equation constrains the allowed radial dependence of the density for a (conformal) analog of the Schwarzschild spacetime. We will return to these consistency conditions below, focusing on the case of the static, spherically symmetric UCO spacetimes considered in this work.

### A. Acoustic geometry for surface gravity waves

As said, in this work, we want to explore the possibility to simulate RBH/UCO geometries and their phenomenology in gravity waves experiments as those performed in Refs. [34,37,38]. This analog system was initially explored in Ref. [39] by considering surface gravity waves in a shallow basin filled with a liquid.[4]

Neglecting viscosity and assuming an irrotational flow, $\mathbf{v} = \nabla\phi$, Bernoulli's equation in the presence of Earth's gravity reads

$$\partial_t \phi + \frac{1}{2}(\nabla\phi)^2 = -\frac{p}{\rho} - gz - V_\parallel + \text{const}, \tag{15}$$

where $\rho$ is the fluid density, $p$ is its pressure, $g$ is the gravitational acceleration, and $V_\parallel$ is an external potential used to sustain a background horizontal flow, denoted $\mathbf{v}_\mathrm{B}^\parallel$.

Once a stationary horizontal background flow is established, small perturbations of the velocity potential, $\delta\phi$, satisfy [39]

$$\partial_t \delta\phi + \mathbf{v}_\mathrm{B}^\parallel \cdot \nabla_\parallel \delta\phi = -\frac{\delta p}{\rho}. \tag{16}$$

Expanding the perturbation potential as a Taylor series in the vertical coordinate,

$$\delta\phi(x, y, z, t) = \sum_{n=0}^{\infty} \frac{z^n}{n!} \delta\phi_n(x, y, t), \tag{17}$$

one can show [39] that, in the long-wavelength regime $\lambda \gg h_\mathrm{B}$, surface waves are well described by the leading component $\delta\phi_0(x, y, t)$. Remarkably, this field propagates as if it were a massless scalar on an effective metric of the form

$$ds^2_\mathrm{AC} = \frac{\rho}{c_\mathrm{sw}}[-(c^2_\mathrm{sw} - \mathbf{v}_\mathrm{B}^{\parallel 2})dt^2 - 2\mathbf{v}_\mathrm{B}^\parallel \cdot d\mathbf{x}dt + d\mathbf{x}\cdot d\mathbf{x}], \tag{18}$$

where $c_\mathrm{sw} \equiv \sqrt{g h_\mathrm{B}}$ is the shallow-water *surface-waves* speed. This is the so-called *acoustic* metric, specialized to the case of surface waves. This construction can be generalized to nonflat bottoms and to background flows with non-negligible vertical components [39]. In the latter case, the generalization is immediate: instead of $\mathbf{v}_\mathrm{B}^\parallel$, one simply finds the entire background velocity $\mathbf{v}_\mathrm{B}$ in the metric. Consequently, by engineering the local depth profile $h_\mathrm{B}(x, y)$ and the background velocity field, one can realize a broad class of effective metrics for surface waves in shallow-water basins.

A comment is in order concerning this metric: in the experiments of our interest, the fluid, often water or superfluid helium, is typically kept in a cylindrical container. We shall then write the metric in cylindrical coordinates

[3]In the BHM limit, models such as the Bardeen or Hayward geometries closely resemble gravastarlike configurations: a large de Sitter core, a thin matter layer, and an exterior that rapidly approaches vacuum at larger radii. The stable inner light ring is usually situated within the thin matter layer and therefore close to the edge of the de Sitter core.

[4]Here, "gravity waves" refers to the fluid-mechanics notion of surface waves whose restoring force is ordinary Newtonian gravity. Waves in the fabric of spacetime are instead called "gravitational waves."

$$ds^2_{\rm AC} = \frac{\rho}{c_{\rm sw}}[-(c^2_{\rm sw} - v^2_r)dt^2 - 2v_r dr\,dt + dr^2 + r^2 d\phi^2 + dz^2], \tag{19}$$

where we have specialized to a purely radial, stationary flow, $\mathbf{v}_{\rm B} = \mathbf{v}^{\parallel}_{\rm B} = v_r \mathbf{e}_r$. While this restriction may appear drastic, it is well motivated for the class of configurations we will consider. The azimuthal velocity is taken zero because we want to simulate a spherically symmetric spacetime. The vertical velocity at the fluid surface plays no role in the acoustic metric for surface waves given in Eq. (18), but we will show for consistency that it is negligible.

Moreover, since we limit our considerations to the surface of the fluid $z = h_B(r)$, $dr^2 + dz^2 = dr^2(1 + h'_B(r)^2)$ in the acoustic metric, where the prime denotes a partial derivative with respect to $r$. This is done not only because treating surface waves is theoretically simpler but also because in the above cited experiments it is only the surface of the fluid that is measured.[5] Furthermore, to allow us to map the acoustic metric to the gravitational metric, which will be written in a form with $g_{rr} = 1$, we shall impose $h'_B(r)^2 \ll 1$. This is *de facto* a dimensional reduction of the metric since it amounts to neglecting the $dz^2$ term. With this, the acoustic metric in cylindrical coordinates has effectively become a $2+1$-dimensional one in polar coordinates.

### B. Correspondence with the target gravitational metric

Now, in order to establish a direct correspondence between the acoustic metric (19) and the gravitational line element (1), it is convenient to rewrite the latter in ingoing Painlevé-Gullstrand (PG) form. This is achieved by the time redefinition

$$dt_{\rm PG} = dt + \frac{\sqrt{2m(r)/r}}{1 - 2m(r)/r} dr, \tag{20}$$

under which the metric (1) becomes

$$ds^2_{\rm GR} = -\left(1 - \frac{2m(r)}{r}\right) dt^2_{\rm PG} + 2\sqrt{\frac{2m(r)}{r}} dr\, dt_{\rm PG} + dr^2 + r^2 d\Omega^2. \tag{21}$$

Here, the subscript "GR" is included to emphasize that this is the target gravitational metric to be simulated. We see, as mentioned above, that $g_{rr} = 1$ in these coordinates. We recall that $d\Omega^2 = r^2 d\theta^2 + r^2 \sin^2\theta d\phi^2$ is the spherical line element. Since this metric is in spherical coordinates, to draw the connection to the $2+1$-dimensional acoustic metric in polar coordinates, we must simply set $\theta = \pi$, placing us on the equatorial plane of the spherically symmetric gravitational metric. Thus, with the $2+1$ polar sonic metric, we are simulating the equatorial plane of the gravitational metric. The fact that the latter is spherically symmetric is the precise reason why realizing only its equatorial plane is enough.

The two metrics clearly look similar now, but there are still two aspects which prevent us from identifying an exact correspondence. First, the acoustic metric has a conformal factor $\rho/c_{\rm sw}$, which we can simply ignore, since it does not affect causal features of the spacetime. Since, eventually, we will be interested in null perturbations when dealing with light rings, echoes, and QNMs, or with inner horizon instabilities (remember also that the surface gravity of a stationary horizon is conformally invariant), the conformal factor will be irrelevant. Second, in the sonic metric, both $c_{\rm sw}(r)$ and $v_r(r)$ are in general functions of $r$. However, we can perform a coordinate transformation in the sonic metric $dt \to dt_{\rm AC} \equiv c_{\rm sw}(r)dt$. Then, the $2+1$-dimensional acoustic metric (dropping henceforth the conformal factor) becomes

$$ds^2_{\rm AC} = -\left(1 - \frac{v_r^2}{c^2_{\rm sw}}\right) dt^2_{\rm AC} - 2\frac{v_r}{c_{\rm sw}} dr\, dt_{\rm AC} + dr^2 + r^2 d\phi^2. \tag{22}$$

Care needs to be taken on the experimental side here because the only time coordinate endowed with an operative meaning is the laboratory time $t$, not $t_{\rm AC}$. This is a manifestation of the lack of diffeomorphism invariance of analog gravity [16]. Using the coordinate transformation $dt \to dt_{\rm AC}$, with $t_{\rm AC}$ standing as the analog of the time coordinate in the gravitational metric is, in a sense, a deliberate desynchronization of clocks, since their ticking is made to depend on their radial position. While this is a nuissance, it is not too problematic, because in surface-wave fluid experiments the sound speed $c_{\rm sw}(r)$ is trackable (in fact, it must be tracked) as a function of $r$. So, by measuring the experimentally accessible laboratory time $t$ and $c_{\rm sw}(r)$, the experimenters can reconstruct and track the desynchronized $t_{\rm AC}(r)$ without any issue.

After these considerations, one can finally make the following identification between fluid and gravitational quantities by imposing a correspondence between the metrics

$$\frac{v_r(r)}{c_{\rm sw}(r)} = -\sqrt{\frac{2m(r)}{r}}. \tag{23}$$

### C. Simulating features of the gravitational spacetime via the fluid velocity profile

As said, depending on the value of the regularization parameter $l$ in $m(r)$ one can have a RBH, an extremal black, hole or a BHM. On the acoustic side, this behavior is evidently controlled by the ratio between the radial velocity and the sound speed, since the appearance of an acoustic horizon is characterized by $v_r(r)/c_{\rm sw}(r) = 1$. Clearly, we expect to find some correspondence between the regularization parameter and this ratio.

[5]This is normally done via a high-resolution camera detecting small differences in the height profile [37,38,40].

To show how through a given velocity profile one simulates the horizon structure of an UCO spacetime, consider the plot in Fig. 2, where the squared ratio $v_r(r)/c_{\rm sw}(r)$ is shown for a horizonless BHM. It exhibits a peak, that is larger/smaller than unity for a RBH/BHM. In this plot, the peak is very close to unity but slightly below, making it a BHM. If the peak were precisely at value unity, we would have two degenerate horizons, making the object an extremal RBH. If it were larger than unity, the curve would cross unity at two specific values of $r = r_\pm$, denoting the outer and the inner horizon of a RBH. Between the two, the region of spacetime is trapped. Outside of the outer horizon the region is obviously untrapped, and, more interestingly, it is untrapped also in the interior of the inner horizon, which is necessary for regularization.

Regardless of which UCO (RBH or BHM) is chosen, $m(r)$ will be a known function of radius. Then, to realize this particular model means that one must find a way to experimentally realize a specific $r$-dependence of $v_r(r)$ and/or $c_{\rm sw}(r)$. In gravity waves fluid experiments, $c_{\rm sw}(r)$ is typically not under direct experimental control, unlike, e.g., in BEC experiments. Hence, it is $v_r(r)$ that needs to be controlled experimentally. However, $v_r(r)$ is not a free function, as it must satisfy the fluid equations and experimental boundary conditions.

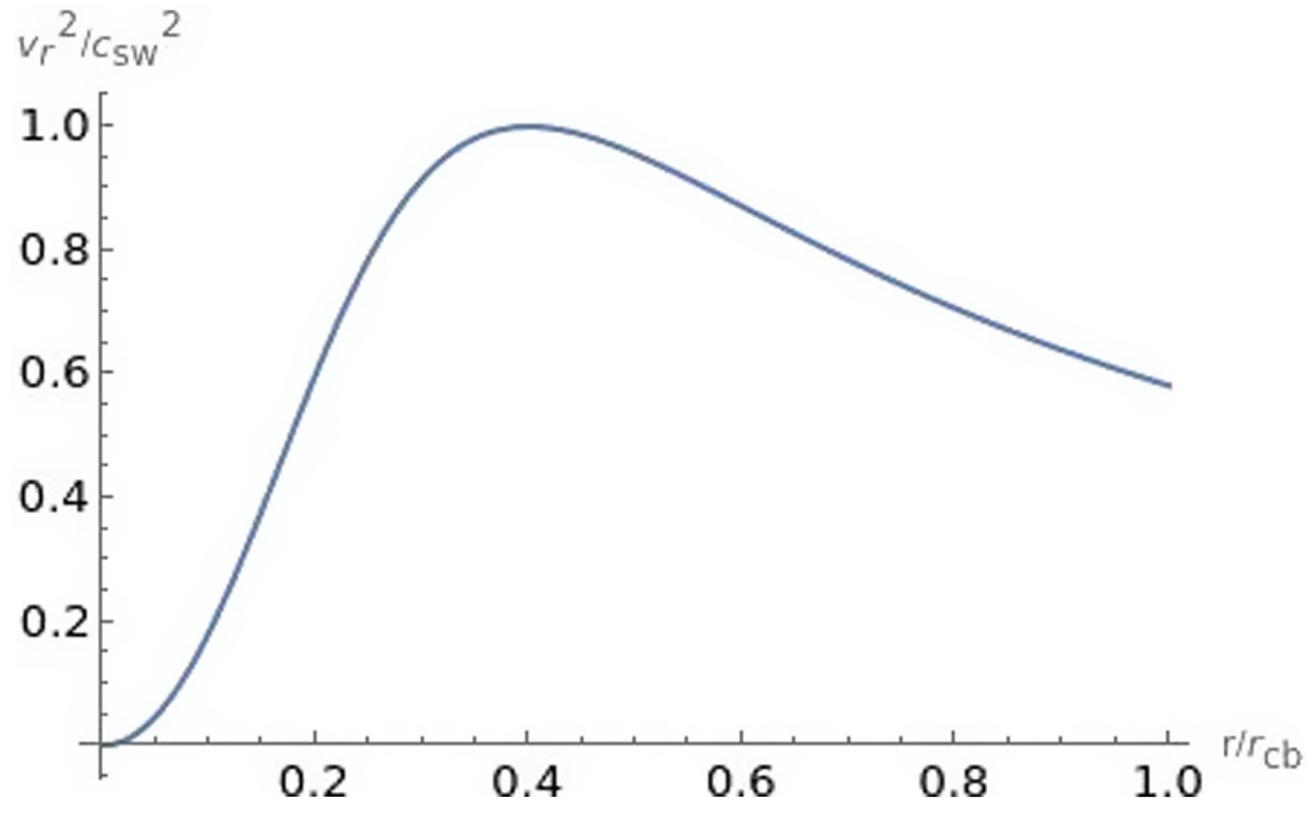

FIG. 2. The desired behavior of $(v_r/c_{\rm sw})^2$ versus radius $r$, normalized to the radius $r_{\rm cb}$ of the container wall, to realize an analog Hayward RBH/BHM. The most important aspect for us is the quadratic behavior close to the origin and the region close to the peak. In this plot, $(v_r/c_{\rm sw})^2$ comes very close to the value 1 but never reaches it, meaning that no horizons form and it is a BHM. If $(v_r/c_{\rm sw})^2$ were strictly bigger than one in a given region, that would mean that it crosses unity twice, corresponding to the outer and the inner horizon of a RBH, with the trapped region sandwiched between them.

The latter are those under experimental control. As we will see in more detail in the next two sections, the experimental setup in fluid experiments is reflected directly in the boundary conditions for the velocity, on the theoretical side. Different experimental setups (for example, a drainage at $r = 0$ or an inflow of fluid at some radius) lead to different boundary conditions for velocity or the fluid height profile. These, in turn, obey differential equations, leading to different functions of $r$ as solutions. So, in practice, this is how one can experimentally exercise control over the function $v_r(r)$ and achieve the desired behavior.

Since $m(r)$ varies from model to model, and since we are mostly interested in the inner region of the spacetime, we shall make a model-independent requirement consisting in prescribing the regularity condition (3), namely, $m(r) \to {\rm const} r^3 + \mathcal{O}(r^4)$ as $r \to 0$, where the arbitrary constant coefficient can be rewritten as $C/l^2$ with $C$ some constant. This is equal to $1/2$ for the Hayward, Dymnikova, and Bardeen metric as well as in other common models.

With this requirement, we get that at the core the interesting acoustic geometries should behave like

$$\frac{v_r(r)}{c_{\rm sw}(r)} \to -\frac{r}{l} \propto r \quad \text{as } r \to 0. \tag{24}$$

This is the correspondence between regularization parameter and the ratio between fluid velocity and sound speed. On the other hand, in the region $r \to \infty$ (or rather $r \to r_{\rm cb}$, where $r_{\rm cb} \gg M$ is the fluid *container boundary*), we would need the behavior

$$\frac{v_r(r)}{c_{\rm sw}(r)} \to -\sqrt{\frac{2M}{r}} \propto r^{-1/2} \quad \text{as } r \to \infty, \tag{25}$$

where $M$ is the constant ADM mass (total mass-energy at infinity) of the spacetime. This is also a model-independent statement, simply expressing the requirement of asymptotic flatness. We will discuss later why we do not need to be rigid about this requirement, although it should be possible to satisfy it.

## V. ANALOG RBH/BHM WITH SURFACE WAVES

In this section, we will gather all formal preliminaries on realizing an RBH/BHM spacetime in analog gravity with surface waves on incompressible fluids. This will allow us, in the subsequent section, to talk about possible experimental realizations.

### A. Fluid equations for surface waves

In this subsection, we collect the equations and mathematical ingredients necessary for our program. The analog system we consider is an irrotational and incompressible fluid, satisfying $\nabla \times \mathbf{v} = 0$ and $\nabla \cdot \mathbf{v} = 0$.

The irrotationality condition is essential in deriving the acoustic metric (25), although extensions of the analog-gravity framework including vorticity have been explored; see, e.g., Refs. [41,42].

The analog system satisfies the Bernoulli equation, the incompressible continuity equation, the irrotationality condition, and a kinematic boundary equation at the surface. In full generality, the above equations are partial differential equations with derivatives with respect to time and space. However, we specialize here to the simulation of a static spacetime and hence assume a stationary fluid flow, meaning that all partial derivatives with respect to time will vanish. In addition, the cylindrical symmetry of fluid analog gravity experiments we are focusing on Refs. [37,38,43–45] presupposes also the vanishing of partial derivatives with respect to the azimuthal angle $\phi$. Furthermore, imposing irrotationality and kinematic boundary conditions at the surface allows us to express $z$-derivatives as $r$-derivatives and therefore, in the end, to deal with just two ordinary differential equations (ODEs) of one independent variable $r$.

The first of these equations is the Bernoulli equation (15), which for a static irrotational flow on the free surface $z = h_B(r)$ without an external potential takes the form [44]

$$\frac{1}{2}\mathbf{v}_B(r)^2 + gh_B(r) - \frac{\sigma}{\rho}\left(h_B''(r) + \frac{h_B'(r)}{r}\right) = \text{const}, \tag{26}$$

where the pressure term at the free surface expresses the surface tension and $\sigma$ is the surface tension coefficient. All other pressure contributions are constant and thus cancel on both sides of the equation. Viscosity has been neglected. We will henceforth suppress the subscript $B$ denoting the background quantities, since we are not exploring wave perturbations in this work.

The constancy of the l.h.s. of this equation implies the equality of its value at the container boundary $r = r_{\text{cb}}$ and at the center $r = r_0 = 0$ of the flow. Therefore, this constant is

$$\text{const} = \frac{1}{2}\mathbf{v}_{\text{cb}}^2 + gh_{\text{cb}} - \frac{\sigma}{\rho}\left(h_{\text{cb}}'' + \frac{h_{\text{cb}}'}{r_{\text{cb}}}\right) = \frac{1}{2}\mathbf{v}_0^2 + gh_0 - \frac{2\sigma}{\rho}h_0''. \tag{27}$$

Please note the absence of the $h'/r$ term on the right-hand side. This is due to the fact that, as a regularity or smoothness condition, the height profile must not slope at the center, hence $h_0' \equiv h'(r=0) = 0$; otherwise, $h'(r)/r$ would diverge there. With this regularity condition, however, it turns out that $\lim_{r\to 0} h'(r)/r = h_0''$, which can be seen as follows. Expand the height profile around the center using the regularity condition $h(r) = h_0 + \frac{1}{2}h_0''r^2 + \mathcal{O}(r^3)$. Then, compute the first derivative of this to see that one obtains precisely $\frac{h'(r)}{r} = h_0''$. This leads to the factor of 2 in the term $\frac{2\sigma}{\rho}h_0''$.

We have not included a quantum pressure term. If the fluid in question is water, there is no quantum pressure; if, instead, we have superfluid helium, the quantum pressure is negligible at hydrodynamic scales since the healing length is nanometer scale [46] (unlike for Bose-Einstein condensates (BECs), which we do not consider here). The surface tension term has already been linearized with respect to $h'(r)$ as required by the possibility to identify the acoustic and gravitational metrics. The velocity at the free surface has been assumed to not depend on $z$; this is quite intuitive, but Appendix A gives more in-depth considerations on this. Note that Ref. [44] imprecisely sets the constant on the right-hand side of the equation to zero. Admittedly, it is irrelevant for their considerations since they look at first-order perturbations of this equation, which eliminates the constant. Since we will be concerned with realizing a particular background spacetime, we will consider this equation at the background level, and the constant will be relevant. It can, indeed, provide a non-negligible contribution to the second derivative of the fluid height profile, as the equation shows. The value of the constant, as made explicit in Eq. (26), depends on the boundary values for velocity, height-profile, and derivatives of the height profile at the boundary of the fluid container ($\mathbf{v}_{\text{cb}}, h_{\text{cb}}, h_{\text{cb}}', h_{\text{cb}}''$) or, equivalently, at the center of the container ($\mathbf{v}_0, h_0, h_0''$). We do not automatically set these to zero, as is often done in the literature, in order to remain general.

The incompressible continuity equation $\nabla \cdot \mathbf{v} = 0$, upon applying the cylindrical symmetry discussed above, reduces to

$$\frac{1}{r}\partial_r(rv_r) + \partial_z v_z = 0. \tag{28}$$

At this stage, no restriction to the free surface has been imposed. Following closely [44], we integrate this equation along $z$, exploiting the assumption that $v_r$ is independent of $z$, to obtain

$$v_z(r,z) = v_z(r, z=0) - \frac{z}{r}\partial_r(rv_r). \tag{29}$$

We now evaluate this expression at the free surface $z = h(r)$. To do so, we invoke the kinematic boundary condition at the free surface, which encodes the physical requirement that general deformations of the surface $h(r,t)$ are generated solely by the normal fluid velocity. Formally, the Lagrangian derivative of the height profile must equal $v_z$ evaluated at the surface [35,39]:

$$v_z(r, z = h(r)) = (\partial_t + \vec{v}\cdot\vec{\nabla})h(r) = v_r h'. \tag{30}$$

This shows that $v_z$ at the free surface is first order in the small quantity $h'$. In the Bernoulli equation, $v_z$ appears only through its square and is therefore negligible to this order. Similarly, in the general acoustic metric for fluids, $v_z$ contributes either quadratically to the $g_{tt}$ component, or through an off-diagonal term proportional to $v_z dz$. Upon restriction to the free surface, the surface element satisfies $dz = h' dr$, so this off-diagonal contribution is likewise $\mathcal{O}(h'^2)$ and can be dropped. This justifies the form of the acoustic line element in Eq. (19), in which both $v_z$ and $dz^2$ have been neglected.

We now insert the above kinematic boundary condition for the vertical velocity at the free surface into Eq. (29), combining $r$-derivatives and renaming $v_z(r, z = 0)$ to $v_{z=0}(r)$ for brevity, obtaining finally

$$\frac{1}{r}\partial_r(hrv_r) = v_{z=0}(r). \tag{31}$$

We see that, even at the free surface $z = h(r)$, the vertical velocity at the bottom of the container $v_{z=0}(r)$ influences the result.

The irrotationality condition is a vector equation which, accounting for axisymmetry, yields three scalar relations:

$$\partial_z v_\phi = 0, \qquad v_\phi = \frac{J}{r}, \qquad \frac{\partial v_r}{\partial z} = \frac{\partial v_z}{\partial r}. \tag{32}$$

We note that $v_\phi$ here carries the dimensions of a linear velocity (length/time), in agreement with Ref. [44], so that the circulation constant $J$ has dimensions of specific angular momentum. By contrast, the authors of Ref. [39] define $v_\phi$ as a proper angular velocity (dimensions 1/time); the two conventions are related by a factor of $r$.

These irrotationality conditions are stated here for completeness. They play no role in the core of this paper, but they do become relevant in Appendizes A and B, where $r$-derivatives are treated more carefully and the assumption of strict spherical symmetry is relaxed to allow for a small but nonzero $v_\phi$. Throughout the main text, we set $L = 0$.

In summary, incorporating all the approximations and symmetry considerations discussed above, the analog system is governed by two fluid ODEs for the variables $v_r(r)$ and $h(r)$: the continuity equation with the kinematic boundary condition substituted in (hereafter "cont. + k.b."), and the Bernoulli equation,

$$\begin{aligned}\frac{1}{r}\frac{d}{dr}(hrv_r) &= v_r h' + h v_r' + \frac{v_r h}{r}\\ &= v_{z=0}(r), \qquad \text{(cont. + k.b.)}\end{aligned} \tag{33}$$

$$\begin{aligned}&h''(r) + \frac{h'(r)}{r} - Q(h(r) - h_0) - \frac{Q}{2g}(v_r(r)^2 - v_0^2)\\ &\quad = 0, \qquad \text{(Bernoulli)},\end{aligned} \tag{34}$$

where $Q \equiv g\rho/\sigma$ and partial derivatives have been replaced by total derivatives since all quantities depend on $r$ alone. The constant $h_0''$ appearing in Eq. (27) has been absorbed into the definition $v_0^2 \equiv v_{0,r}^2 - 4gh_0''/Q$ for compactness. Throughout, subscript-0 quantities denote values at the origin $r = 0$ rather than at the container boundary, reflecting our choice to integrate the ODEs outward from the center.

Before proceeding, it is worth pausing to give a physical interpretation of $Q \equiv g\rho/\sigma$. Its inverse square root defines the capillary length $l_c \equiv Q^{-1/2} = \sqrt{\sigma/(\rho g)}$, which is the characteristic scale at which gravitational and surface-tension forces are comparable. For length scales $L \gg l_c$ gravity dominates, while for $L \ll l_c$, surface tension dominates. Equivalently, the dimensionless combination $\mathcal{Q} = Qr_{\rm cb}^2 = (r_{\rm cb}/l_c)^2$ is nothing other than the Bond number of the system, quantifying the relative importance of gravity over surface tension at the scale of the container. In typical experimental setups we are interested in here, large values of $\mathcal{Q}$ are expected; for both water and superfluid helium at centimeter-to-meter scales, gravitational forces overwhelmingly dominate surface tension.

### B. Boundary values and conditions

The source term $v_{z=0}(r)$ in Eq. (33) and the boundary velocity $\vec{v}_0$ in Eq. (34) are both determined by the experimental setup and thus under direct control. The central height $h_0$ is under indirect experimental control: given $v_0$, it is fixed by evaluating the Bernoulli equation at $r = 0$. To solve the continuity equation, the radial component $v_{0,r}$ must be specified independently, while $h_0' = 0$ is already given as a regularity condition.

Let us illustrate how the experimental setup reflects the form of the velocities $v_{z=0}(r)$ and $v_{0,r}$, using the former as an example. Most experimental setups rely on some drainage mechanism at the center. In this case, this boundary velocity[6] will be some negative function of $r$, often approximated as constant, in a radius $d$ from the center and zero at all other points of the bottom of the container: $v_{z=0}(r) = -U\Theta(d - r)$ [44] with positive constant $U$. Other authors restrict their considerations only to the area of the container away from the drain, and so automatically set $v_{z=0}(r)$ to zero [35,39]. This is where the solution of the (cont. + k.b.) equation (33) $v_r(r) = \frac{\rm const}{rh(r)}$ commonly found in the literature comes from. We do not impose $v_{z=0}(r) = 0$, remaining general in order to leave room for novel experimental setups that could be better suited for realizing an UCO background spacetime. The importance of this will become evident in Sec. VI, where we will propose concrete analog experimental realizations of the core and the asymptotic region of the spacetime.

[6]This is not a boundary condition in the usual sense in the context of differential equations, but it is a velocity at a physical boundary.

The structure of the mathematical problem is then transparent: the experimentally imposed velocity boundary conditions and boundary values constitute the mathematical encoding of the physical setup. Given these, the (cont. + k.b.) equation (33) determines the radial velocity profile $v_r(r)$, which in turn enters the Bernoulli equation (34) to yield the free-surface height profile $h(r)$.

### C. Speed of surface waves

To establish the correspondence between the sonic and gravitational metrics, one final equation must be satisfied: the identification equation (23) relating fluid and gravitational quantities. The third condition we impose on the fluid system is therefore

$$v_r(r) = -\sqrt{\frac{2m(r)}{r}\tilde{g}(r)h(r)}, \tag{35}$$

where $\tilde{g}(r)$ is an effective gravitational acceleration that depends in general on $r$, receiving contributions from $v_r(r)$, $h(r)$, and their derivatives [35,39]. Whether it is possible to simultaneously satisfy the fluid equations, which are dictated by nature, and this identification equation, which we wish to impose in order to realize a UCO spacetime, requires careful analysis; this is the subject of Sec. V D. We first turn to determining $\tilde{g}(r)$.

In the shallow-water regime, where the wavelength of perturbations greatly exceeds the fluid height, the wave speed is given by $c_{\rm sw}^2 = gh$ [16], as has been mentioned. It turns out that using the constant $g$ in this formula is actually a simplification. Indeed, when the background fluid height is nonuniform, fluid elements traveling along the surface experience additional forces beyond the purely vertical gravitational one. This gives rise to a position-dependent effective gravitational acceleration (EGA) $\tilde{g}(r)$, and hence a spatially varying wave speed $c_{\rm sw}^2(r) = \tilde{g}(r)h(r)$. Following Ref. [35], the EGA reads explicitly

$$\tilde{g}(r) = g + (\partial_t + \vec{v}\cdot\vec{\nabla})^2 h(r) = g + v_r^2 h'' + v_r v_r' h', \tag{36}$$

where $(\partial_t + \vec{v}\cdot\vec{\nabla})^2 \equiv \frac{D^2}{Dt^2}$ is the second Lagrangian derivative. The correction term thus represents the acceleration of the height profile as experienced by a fluid element comoving with the flow and can be interpreted as the fictitious forces felt by this noninertial observer. This point is elaborated in Appendix C, where a more detailed illustration and worked example are provided.

With this, the identification equation (35) becomes a complicated nonlinear ODE,

$$v_r^2 = \frac{2m(r)}{r}(gh + v_r^2 h'' h + v_r v_r' h' h), \tag{37}$$

where we have made explicit only the $r$-dependence of $m(r)$ to remind ourselves that this is a known function of $r$ and is not an extra variable that is added besides $v_r(r)$ and $h(r)$.

Unless $h(r)$ is well approximated by a constant, it is not immediately obvious that the derivative corrections to $\tilde{g}(r)$ are negligible. The second extra term in the EGA is linear in $h'$, so unless $v_r'$ itself carries at least a linear dependence on $h'$, it cannot be neglected. To assess the order of magnitude of the term involving $h''$, we draw an analogy with the slow-roll conditions of inflation. For the identification between the sonic and gravitational metrics to hold, $h'$ must not merely be small at a single point but must remain small throughout the container, i.e., over the entire range $r \in [0, r_{\rm cb}]$. This implies that the dimensionless combination $h''h$ must also be much smaller than unity. However, this does not entail that it is negligible compared to $h'$: in slow-roll inflation, the dimensionless second derivative of the relevant field can be of the same order as, or even larger than, the dimensionless first derivative, while still constituting a small correction. By the same token, the $h''$ term in the EGA cannot in general be discarded. Nevertheless, as we argue in the following subsection, neglecting both derivative terms turns out to be necessary to make the ODE system tractable, for reasons rooted in its structure.

### D. Imposing the identification equation

Suppose the (cont. + k.b.) equation (33) has been solved for $v_r(r)$, using the experimental inputs $v_{z=0}(r)$ and $v_r(0) = v_{0,r}$. The solution $v_r(r, h(r))$ is expressed in terms of $r$ and the still-unknown function $h(r)$. Substituting this into the Bernoulli equation yields one ODE for $h(r)$. Simultaneously, the identification equation (37) furnishes a second ODE for $h(r)$ upon the same substitution. The requirement of realizing a RBH/BHM spacetime therefore imposes an additional, potentially nonlinear second-order ODE on $h(r)$ beyond the one already dictated by hydrodynamics.

In general, this constitutes an overdetermined system of ODEs [47], which admits no common solution unless the two equations are equivalent or one is a derivative of the other; neither of which holds generically. The only viable strategy is to enforce their equivalence by imposing a further constraint. Writing both equations with $h''$ isolated on the left-hand side,

$$h'' = Q(h(r) - h_0) + \frac{Q}{2g}(v_r(r, h(r))^2 - v_0^2) - \frac{h'(r)}{r} \equiv A(r), \tag{38a}$$

$$h'' = \frac{r}{2m(r)h(r)} - \frac{g}{v_r^2(r, h(r))} - \frac{h'(r)v_r'(r, h(r))}{v_r(r, h(r))} \equiv B(r), \tag{38b}$$

the two ODEs admit a common solution only if the reduced constraint $A(r) = B(r)$ is satisfied. This is equivalent to eliminating $h''$ by substituting one equation into the other, leaving a first-order ODE to be solved. This condition is necessary but not sufficient; any solution must subsequently be inserted back into the original second-order equations to verify consistency, since second-order ODEs require two integration constants rather than one. A simple worked example is given in Appendix D.

The reduced constraint equation reads explicitly

$$Q(h(r)-h_0)+\frac{Q}{2g}(v_r(r,h(r))^2-v_0^2)-\frac{h'(r)}{r} = \frac{r}{2m(r)h(r)}-\frac{g}{v_r^2(r,h(r))}-\frac{h'(r)v_r'(r,h(r))}{v_r(r,h(r))}. \quad (39)$$

This is a complicated, nonlinear first-order ODE whose solutions generally cannot be expressed in closed form, and which must additionally be checked for consistency against the original second-order equations.

While this program can be carried out numerically, the problem simplifies considerably if $h(r)$ is well approximated by a constant, allowing both $h''$ and $h'/r$ to be safely neglected. The latter approximation is supported by two additional arguments in our regimes of interest. As $r\to 0$, we have $h'(r)/r\to h_0''$, so if $h''$ is negligible, so is $h'(r)/r$. As $r\to\infty$, this term decays in any case [44]. In physical terms, neglecting both surface-tension corrections amounts to requiring that $h(r)$ varies on scales much larger than the capillary length $l_c\equiv Q^{-1/2}$. Defining the characteristic length scale of variation as $L\sim h/h'\sim\sqrt{h/h''}$, this condition is simply $L\gg l_c$ as we already saw.

Under this approximation, the sound speed reduces to $c_s=\sqrt{gh}\approx\text{const}$, and all equations except the (cont. + k.b.) equation (33) become algebraic, eliminating the problem of ODE overdetermination. Another, experimental, argument for the necessity of seeking setups with approximately constant $h(r)$ is provided in Sec. VI A. An alternative (but more complicated) way to resolve this overdetermination is to introduce a non-flat container bottom, as outlined in the Appendix E.

## VI. EXPERIMENTAL CONSIDERATIONS

In this section, we propose specific experimental setups for realising the spherically symmetric RBH/BHM spacetime of interest. We begin by complementing the mathematical argument of the previous subsection with an experimental one, showing that setups with negligible $h'$ and $h''$ are necessary to remain within the regime of validity of the analog gravity framework.

### A. Numerical estimates

By inserting numerical values for the relevant parameters in surface-gravity wave setups based on superfluid helium or water, we will demonstrate that a valid analog gravity system requires an approximately constant fluid height, $h(r)\approx h_0$, near the center.

We begin by restating the Bernoulli equation with $h''$ isolated:

$$h''(r)=Q(h(r)-h_0)+\frac{Q}{2g}(v_r(r)^2-\vec{v}_0^2)-\frac{h'(r)}{r}. \quad (40)$$

Whether $h''$ is large or small in absolute value is thus controlled by the right-hand side. Working near the center, we approximate $h'(r)/r\approx h_0''$ as before and absorb it into the constant $v_0^2$.

Since $h'$ is assumed small, neither $(h(r)-h_0)$ nor $(v_r(r)^2-v_0^2)$ can be large in absolute value, the latter because it is indirectly tied to $h'$ through the (cont. + k.b.) equation (33) and because large velocity gradients would render the experiment unstable. The dominant factor controlling the magnitude of $h''$ is therefore $Q$.

To make this precise, we render the equation dimensionless. The problem naturally provides the capillary length $l_c=Q^{-1/2}$ as the characteristic length scale, but since this is precisely the quantity whose magnitude we wish to assess, we instead use the container boundary radius $r_{\rm cb}$ as the reference scale. Let us then define the dimensionless variables

$$\mathcal{R}\equiv\frac{r}{r_{\rm cb}},\qquad \mathcal{H}\equiv\frac{h}{r_{\rm cb}},\qquad \frac{\mathcal{V}_r^2}{g}\equiv\frac{v_r^2}{gr_{\rm cb}},\qquad \mathcal{Q}\equiv Qr_{\rm cb}^2, \quad (41)$$

in terms of which the Bernoulli equation, multiplied through by $r_B$, becomes

$$\mathcal{H}''(\mathcal{R})=\mathcal{Q}(\mathcal{H}(\mathcal{R})-\mathcal{H}_0)+\frac{\mathcal{Q}}{2g}(\mathcal{V}_r(\mathcal{R})^2-\mathcal{V}_0^2), \quad (42)$$

where primes now denote derivatives with respect to $\mathcal{R}$.

For a superfluid helium experiment [37], with $Q\approx 4.35\ \text{mm}^{-2}$ and $r_B=37.3$ mm, one obtains $\mathcal{Q}\approx 6064$. Taking as a reference the dimensionless height at the container wall, $\mathcal{H}_{\rm cb}\equiv h_{\rm cb}/r_{\rm cb}=15\ \text{mm}/37.3\ \text{mm}\approx 0.4$, the value of $\mathcal{Q}$ is seen to be vastly larger, so $\mathcal{H}''$ would be enormous, and the "slow-roll" condition $\mathcal{H}''\ll 1$ would be completely violated. For a typical water experiment [40,43] using a flume of length/width $r_{\rm cb}\sim 1$ m and comparable depth, with $Q\approx 1.36\times 10^5\ \text{m}^{-2}$, one finds $\mathcal{Q}\approx 13600$, an even more unfavorable value.

The only viable route is therefore to find an experimental setup in which both the height deviation $(\mathcal{H}(\mathcal{R})-\mathcal{H}_0)$ and the velocity contrast $(\mathcal{V}_r(\mathcal{R})^2-\mathcal{V}_0^2)/g$ are so small that, despite the large value of $\mathcal{Q}$, the resulting $\mathcal{H}''$ remains small and negligible. We also note that $\mathcal{Q}\propto r_{\rm cb}^2$ can be reduced by choosing a smaller container. However, the container must simultaneously be large enough to accommodate both the core of the simulated compact object and a sufficiently extensive exterior region, effectively an asymptotically flat zone, where quasinormal modes can be measured.

### B. Central drainage setup

Having established on multiple grounds the necessity of finding setups with approximately constant $h(r)$, we now examine whether this requirement can be met in the typical experimental configuration used for surface-gravity wave analogs: one with a central drainage mechanism [34,37,38,40,44]. A key distinction from these references is that we do not consider rotation.

With $h''$, $h'$, and $v_z = v_r h'$ consistently neglected everywhere (including inside $v_0^2 = v_{0,r}^2 - \frac{4g}{Q} h_0''$), the governing equations reduce to

$$\frac{1}{r}(hrv_r)' = v_{z=0}(r), \qquad (\text{cont.} + \text{k.b.}) \tag{43}$$

$$2g(h(r) - h_0) - (v_r(r)^2 - v_{r,0}^2) = 0.(\text{Bernoulli}). \tag{44}$$

We model the drainage as imposing a spatially uniform downward velocity over a disk of radius $d$ centered at the origin,

$$v_{z=0}(r) = -U\Theta(d - r), \tag{45}$$

where $U > 0$ is a constant and $\Theta$ is the Heaviside step function. Solving the (cont. + k.b.) equation (33) with this source yields [44]

$$v_r(r) = \begin{cases} -\frac{Ur}{2h(r)}, & r < d, \\ -\frac{Ud^2}{2rh(r)}, & d \le r \le r_{\text{cb}}. \end{cases} \tag{46}$$

For the ratio $v_r/c_{\text{sw}}$ to behave as $-r/l$ near the origin (as required by the analog spacetime construction), $h(r)$ must be approximately constant there, making $c_{\text{sw}}$ constant as well. A strictly constant $h(r) = h_0$ is not viable, since the Bernoulli equation would then force $v_r = v_{r,0}$ everywhere; but working in the approximation $h(r) \approx h_0$ and retaining the velocity profile above is a good approximation, provided the Bernoulli solution for $h(r)$ is sufficiently flat in the central region.

By symmetry, fluid flows toward $r = 0$ from both sides, so the only physically admissible value is $v_{r,0} = 0$. The Bernoulli equation then reduces to the cubic

$$h^3 - h_0 h^2 + \frac{U^2 r^2}{8g} = 0, \tag{47}$$

whose unique positive real solution is plotted in Fig. 3.

Near the origin the solution is well approximated by a constant, confirming that the desired behavior is at least approximately achievable. A noteworthy discrepancy, however, is that once the approximation breaks down the height profile is concave, whereas experiments show a convex profile near the drain (see Fig. 1(a) of Ref. [38] for a photograph of the actual fluid surface). This is most likely a consequence of the rotation present in that experiment: the drainage in current superfluid helium setups is realized via a draining vortex, making the analog system one of a rotating, axisymmetric object rather than a spherically symmetric one. In Appendix B, we explore the inclusion of slow rotation in both the gravitational and acoustic metrics, with a view to simulating a slowly rotating UCO with mass function $m(r)$ and small rotation parameter $a$, but encounter difficulties in identifying the rotational fluid and gravitational degrees of freedom. For this reason, and because a spherically symmetric UCO appears simpler to realize on theoretical grounds, we advocate using water rather than superfluid helium.

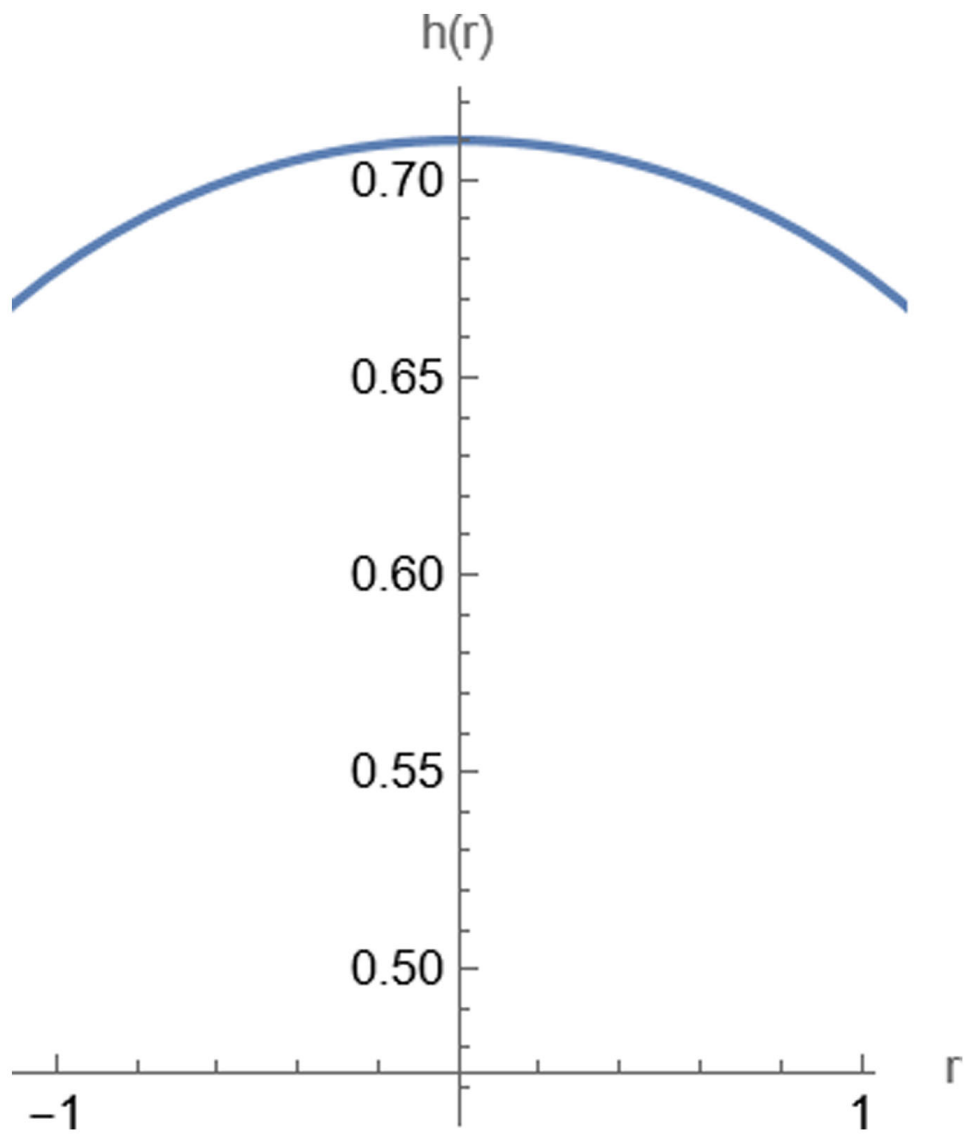

FIG. 3. Positive real solution $h(r)$ of the cubic Bernoulli equation (47) for the central drainage setup with constant $v_{z=0}$. Units for $h$ and $r$ are arbitrary. The solution should be matched to the exterior profile at $r = d$; from the plot, $d \approx 0.2$ appears to be a good choice for remaining well within the approximately constant regime.

A further source of discrepancy is that a physical drainage system does not produce a perfectly uniform $v_{z=0}(r)$. In practice, the region $r \le d$ is often excluded from the analysis precisely because $h(r)$ drops sharply there, invalidating both the shallow-water approximation and the analog gravity construction. That said, stabilizing the central region (the bottom of the dip) is not in principle impossible, but this is a challenge of exclusive experimental nature on which we cannot say more here.

Assuming such stabilization is achievable, we can identify the regularization parameter of an analog BHM spacetime and estimate the position of the inner light ring, where the accumulation of long-lived modes responsible for the associated instability is expected to occur. Inserting

$v_r = -Ur/(2h_0)$ and $c_{\rm sw} = \sqrt{gh_0}$ into the identification equation (24) gives

$$l = \frac{2h_0}{U}\sqrt{gh_0}. \tag{48}$$

This regularization length is set by the central drainage velocity and the central fluid height. For a near-extremal UCO, one close to horizon formation, the condition $l \gtrsim l_* \equiv \sqrt{16/27}M$ provides an estimate of the ADM mass of the simulated spacetime and hence of the inner light ring position $r_{\rm ILR} \sim M$ (noting that the precise numerical coefficient is model dependent [8,48,49]),

$$r_{\rm ILR} \sim \sqrt{\frac{27}{16}}l \sim \frac{3h_0}{2U}\sqrt{3gh_0} \sim 3\ {\rm m}, \tag{49}$$

where the numerical value is obtained with $h_0 = 0.5$ m and $U = 1\ {\rm m\,s^{-1}}$.

This scale is completely out of reach for superfluid helium experiments, whose containers are typically ~10 mm across. In water it is more feasible in principle, but 3 m is still large even for the central drain region $r \leq d$. One could increase $U$, though this risks destabilising the flow. A more practical alternative is to reduce $h_0$: for $h_0 = 0.1$ m, one obtains the more manageable $r_{\rm ILR} \sim 0.25$ m, with corresponding $l \sim 0.2$ m and $M \sim 0.25$ m. This could be realized in a water container of a few meters in diameter with a central drain of width $d \gtrsim 0.3$ m.

Further numerical estimates of potential limiting factors to the experimental realization like the onset of turbulence, of significant viscous dissipation, or of nonlinear dispersion, are given in Appendix H.

Finally, we examine whether the exterior solution ($d \leq r \leq r_{\rm cb}$) can reproduce the asymptotic behavior at large $r$ given in Eq. (25), for sufficiently large $r_{\rm cb}$. Since this solution coincides with that of Appendix A of Ref. [44], where a power-series expansion including $h(r)$ is provided, we find that the required asymptotic behavior is not reproduced. This motivates us to consider alternative experimental configurations in the following subsection.

### C. Global gradient drainage

To realize the identification equation (23) with the power-law asymptotic behaviors (24) and (25), the two quantities at our disposal are $v_{z=0}(r)$ and $v_r(r=b)$. Now, $b$ denotes a chosen boundary point beyond which the fluid can be considered to be in the $r \to \infty$ regime. To make this dependence on the boundary velocities explicit, we integrate the generic (cont. + k.b.) equation (31) to obtain

$$v_r(r) = \frac{1}{rh(r)}\int rv_{z=0}(r)dr + \frac{B}{rh(r)}, \tag{50}$$

where the integration constant $B$ is fixed by the boundary value $v_r(b)$:

$$B = bh(b)v_r(b) + \int_0^b rv_{z=0}(r)dr. \tag{51}$$

With $v_r(r)$ thus uniquely determined by $v_{z=0}(r)$ and $v_r(b)$, the Bernoulli equation—algebraic when surface tension is neglected—yields $h(r)$. The ratio $v_r/c_{\rm sw} = v_r/\sqrt{gh(r)}$ entering the identification equation is therefore also uniquely determined by these two quantities, which are precisely the experimentally tunable inputs available to simulate a desired mass profile $m(r)$, as was elaborated in less general terms in Sec. V B.

In the experimental context, a general (nonconstant) negative $v_{z=0}(r)$ corresponds to a spatially varying drainage distributed across the entire container floor, what we term a *global gradient drainage*. In practice, this could be realized in discretized form by partitioning the floor into concentric annular segments, each with a different drainage strength, approximating a smooth $r$-dependence of $v_{z=0}(r)$ in the continuum limit.

Specifically, we examine whether the asymptotic behavior $v_r/c_{\rm sw} \propto r^{-1/2}$ as $r \to \infty$ can be reproduced by a power-law drainage profile $v_{z=0}(r) \propto r^a$. The restriction to power laws is well motivated; since $v_r(r)$ is obtained by integrating $v_{z=0}(r)$ and $h(r)$ by taking a square root of $v_r(r)$ via the Bernoulli equation, a power-law source is the natural, and essentially unique, way to generate a power-law ratio $v_r/\sqrt{gh(r)}$ asymptotically.[7] Power laws are also among the most straightforward profiles to implement experimentally.

Thus, we model the global gradient drainage in the two asymptotic regimes as

$$v_{z=0}(r) = \begin{cases} -U, & r \to 0, \\ \cdots & \\ -W(\frac{r}{b})^a, & r \to \infty, \end{cases} \tag{52}$$

retaining the constant-$U$ behavior at the center, which we have already shown to be compatible with our requirements, and leaving the intermediate region unspecified. Here, $W > 0$ is a constant setting the drainage strength at $r = b$, the outer boundary of the intermediate region, at which a boundary value for the radial velocity is imposed. This boundary value is determined by matching the velocity continuously across $r = b$, i.e., by requiring $v_r(b^-) = v_r(b^+)$. The velocity at $b$ in turn depends on the velocity at the outer edge $d$ of the central region, which is itself fixed by the condition $v_{r,0} = 0$. The overall procedure is therefore a stepwise outward integration;

[7] More general functional forms will generically reduce to a power law in their asymptotic expansion.

starting from $v_{r,0} = 0$ at the origin, the radial velocity is calculated region by region toward larger r, with continuity enforced at each boundary.

Focusing on the asymptotic region $r \to \infty$,[8] Eq. (50) gives the radial velocity

$$v_r(r) = -\frac{W r^{a+1}}{(a+2) b^a h(r)} + \frac{B}{r h(r)}. \tag{53}$$

Substituting this into the Bernoulli equation (44) and distinguishing the three cases $a \gtrless -1$ and $a = -1$, we analyze the leading terms as $r \to \infty$, retaining only the dominant contributions. This reduces the Bernoulli equation to a tractable form, from which the leading behavior of $h(r)$ follows by straightforward algebra:

$$h(r) \sim \begin{cases} \text{no positive solution}, & a > -1, \\ h_{\rm cb}, \text{determined by a cubic equation}, & a = -1, \\ h_{\rm cb}, \text{determined by a linear equation}, & a < -1. \end{cases} \tag{54}$$

For $a > -1$, the structure of the Bernoulli equation precludes any positive solution, indicating that the fluid surface has reached the bottom of the container and cannot be continued beyond. The algebra and the explanation of this case is shown in Appendix F. In the remaining two cases, the leading-order height is an $a$-independent constant $h_{\rm cb}$, since the Bernoulli equation is dominated by constant terms in both limits.

With the above result for $h(r)$, we obtain for the l.h.s. of the identification equation,

$$\frac{v_r}{\sqrt{g h(r)}} \sim \begin{cases} \text{undefined}, & a > -1 \\ -\frac{W}{b^a h_{\rm cb}^{3/2}} & a = -1 \\ -\frac{W}{b^a h_{\rm cb}^{3/2}} r^{a+1} - \frac{B}{h_{\rm cb}^{3/2} r} & a < -1 \end{cases}, \tag{55}$$

where in the first case the velocity is undefined since the fluid is no longer present in the asymptotic region, $h(r)$ having reached zero. We observe that we obtain a non-constant power-law only in the case $a < -1$. Which of the two terms is the leading one depends on a further case distinction $a \lesseqgtr -2$. We can obtain the desired behavior $r^{-1/2}$ for the value $a = -3/2$, i.e., for $v_{z=0}(r) \to -W(\frac{r}{b})^{-3/2}$ as $r \to \infty$. Comparing the expression in the above equation (55) with the corresponding gravitational equivalent in Eq. (25), we identify

$$M = \frac{W^2 b^3}{2 h_{\rm cb}^3}. \tag{56}$$

We can then see that the analog ADM mass of our spacetime is given by the draining velocity at the point $r = b$, i.e., $W = v_{z=0}(r = b)$, and by the value of that point and by the asymptotic fluid height. The directly tunable experimental quantity among these three is $W$, which allows for the ADM mass of the simulated spacetime to be adjusted at will. Let us note that this direct experimental control of $M$ could allow us to generalize the class of spacetimes we can simulate even further, which we briefly elaborate on in Appendix G.

We have checked, for consistency, whether including the $h''$ or $h'$ terms in both the Bernoulli equation and the wave velocity $c_{\rm sw}(r)$ significantly changes these results. We find that those terms are consistently negligible, as we required in the previous sections. We also checked whether changing the drain into an inflow (changing the sign in front of $W$), like a hydrojet in a pool, could change this, but of course the exponent of $r$ is not sensitive to the sign of $W$.[9]

We have thus shown that, in principle, also the asymptotic region of the spacetime in question should be realizable experimentally if a gradient drain can be implemented. One main difficulty in doing this is to find a way to refill the fluid container without excessively perturbing the system, since the global drainage will quickly lead to the loss of the fluid. In current rotating fluid experiments like Refs. [34,37,38], the refilling mechanism is exactly at the boundary of the container, meaning that $v_{z=0}(r)$ has the wrong sign there, for our purposes.

If the refilling complicates the realization of the asymptotic region, let it be noted that it is actually not as essential as the central region. It is true that the QNM/echo signal is observed at infinity, but it is "generated" in the inner region of the spacetime. This is why the observation of the signal is not impeded by not precisely realizing the asymptotic region, if the curvature drops to zero there. Fortunately, it does, as can be checked by substituting the results in Eq. (55) as $-\sqrt{2m(r)/r}$ into the Kretschmann scalar in Eq. (2).

Let us make one concluding remark. In this subsection, we have provided the unique $v_r(r)$ profile (53) and $h(r)$ profile (54), self-consistent solutions of the fluid equations in the asymptotic region $r \to \infty$, that provide an exact correspondence between the gravitational metric (21) and the acoustic metric (22) in that region. This means that $m(r) \to M$ is reproduced *exactly* by these fluid quantity profiles.[10] Similarly, in the Sec. VI B, up to the approximation of a constant $h(r) = h_0$, the correspondence between fluid and gravitational quantities in the central region $r \to 0$ is exact. However, we do not attempt a smooth joining between these two limiting regions, because realizing them is enough, and realizing the entire Hayward model $m(r) = \frac{Mr^3}{r^3+2Ml^2}$ would be unnecessarily complicated.

[8]Recall that, in a finite container of radius $r_{\rm cb}$, this amounts to taking $r_{\rm cb} \to \infty$ in units of $M$, and then sending $r \to r_{\rm cb}$.

[9]Obviously, the sign of $v_r$ is sensitive to the sign of $W$, and a positive sign in front of $W$ will lead to positive $v_r$ regions, which is not viable for a UCO spacetime.

[10]Note that if these profiles are realized in the entire fluid container, one would have an analog of the Schwarzschild black hole, for which the equality $m(r) = M$ is valid for all $r$.

## VII. CONCLUSIONS

Efforts in analog gravity have thus far been directed primarily toward realizing a horizon or ergoregion and observing their signatures in the form of analog Hawking radiation or superradiance. The present work advances this program toward the simulation of a concrete spacetime in its entirety, rather than merely its horizon or ergoregion structure, with the goal of accessing analog signatures associated with light rings as well. Focusing on static, spherically symmetric spacetimes with mass function $m(r,l)$, we have reviewed how the regularization parameter $l$ eliminates the central singularity, making the central object either a RBH or a horizonless BHM. Which of the two is realized depends crucially on $l$, which controls the finiteness of curvature scalars as $r \to 0$, determines the horizon structure, and governs the spacetime geometry near the core. In particular, $l$ sets the compactness and thereby the time delay between echoes—observable if the object is a UCO and potentially carrying signatures of a conjectured dynamical instability of the inner light ring. QNMs carry related signatures. For a RBH, the inner horizon instability originates from behind the horizon and cannot be read off directly from the QNM spectrum; however, if a new equilibrium state is established, the nature of that final object can in principle be inferred from its QNM spectrum. The overarching motivation of this work is therefore to test, in an analog gravity laboratory setting, whether the analog object suffers the inner horizon/inner light ring instability and what the resulting final state is. Although the analog system obeys different dynamical equations, we have identified theory-independent, kinematical features of the signal from which general conclusions can be drawn.

Our analog system is the free surface of an incompressible, irrotational fluid. Surface-gravity waves on this surface satisfy a wave equation whose d'Alembertian is computed with respect to an effective acoustic metric, which underpins the analogy. Comparing the acoustic and gravitational line elements, we have shown that an exact correspondence between fluid and gravitational quantities exists in the form $v_r/c_{\rm sw} = -\sqrt{2m(r,l)/r}$, provided the laboratory clocks are locally desynchronized according to $dt \to dt_{\rm AC} = c_{\rm sw}(r)dt$—a correction that must be tracked experimentally. We have formulated the continuity, Bernoulli, and kinematic boundary equations at the fluid surface and provided a detailed analysisboth mathematical and experimentalof why derivative terms $h'$ and $h''$ in the fluid height should be neglected throughout: both in the fluid equations and in the sound speed $c_{\rm sw}(r) = \sqrt{\tilde{g}(r)h(r)}$. Such derivative terms were shown to appear the latter through fictitious acceleration contributions to an effective gravitational acceleration $\tilde{g}(r)$.

We have emphasized that the experimental setup enters the equations through the boundary velocity $v_0$ at the center and the externally controlled bottom velocity profile $v_{z=0}(r)$. Many theoretical treatments set these to zero, explicitly or implicitly; we have kept them general in order to make their role transparent in simulating concrete spacetimes. As an illustration, we have analyzed a nonrotating central drainage setup and shown that, under the approximation of constant $h(r)$, the inner region of a static spherically symmetric UCO/RBH spacetime can be simulated. With an efficient and noninvasive refilling mechanism, the exterior region can also be reproduced via a gradient drainage, though this is not strictly necessary for measuring echoes or QNMs as long as the curvature remains finite there—which it does. The central and asymptotic regions together yield analog expressions for the two key gravitational parameters: the regularization parameter and the ADM mass,

$$l = \frac{2h_0}{U}\sqrt{gh_0}, \qquad M = \frac{W^2 b^3}{2h_{\rm cb}^3}, \tag{57}$$

which are ultimately set by the drainage velocities $U$ (central) and $W$ (at intermediate radius $b$). These same velocities determine the asymptotic fluid height $h_{\rm cb}$ through the Bernoulli equation, while the central height $h_0$ is a boundary value given by the setup. The gravitational quantities $l$ and $M$ are therefore tunable through $U$, $W$, and $h_0$.

A large part of the difficulty in realizing an exact spacetime mapping with surface-gravity waves can be traced to the mismatch between the number of fluid variables and the number of equations. Imposing the identification equation adds an extra constraint on top of the fluid equations for $(v_r(r), h(r))$, making the system overdetermined. With experimental control limited to boundary values or asymptotic behaviors of the velocity, realizing the mapping for a given model $m(r,l)$ is generally only possible under approximations of varying justifiability.

One way to resolve this overdetermination, which we have outlined in the Appendix E but defer to future work, is to introduce an additional variable: the curvature $f(r)$ of a nonflat container floor. This quantity enters both the hydrodynamic equations and the identification equation and is directly experimentally controllable, unlike the velocity profile which is only indirectly accessible through its boundary behavior. Adding an extra experimentally controlled variable to the system of three ODEs eliminates the problem of ODE overdetermination and, in principle, makes it possible to simulate any target static, spherically symmetric spacetime specified by $m(r)$.

An even more promising approach is that of a Bose-Einstein condensate as an analog system. There, instead of the variable $h(r)$, there are three other variables besides the radial velocity $v_r(r)$: the density $\rho(r)$ since the system is compressible, an external potential $V_{\rm ext}(r)$ that is used to control the system experimentally, and the sound speed $c_s(r)$ that is under experimental control via the Feshbach resonance [36,50–52]. The total number of dependent variables entering the hydrodynamical equations is thus 4 $(v_r, \rho, V_{\rm ext}, c_s)$, and there are three ODEs: the continuity, Bernoulli and identification equations. Therefore, with BECs, the problem of ODE overdetermination is totally circumvented. This simple "variable-counting" argument illustrates the considerably greater potential for BEC analog systems for our purposes with respect to the present surface-gravity fluid systems, and we intend to explore this potential in future work.

Finally, we stress that even in the surface-gravity wave case already known simple models for $m(r,l)$ may prove too complex to realize experimentally in full detail. For the purposes of measuring QNMs or echoes, the essential requirements are that curvature is finite and that $m(r,l)$ grows at least as $r^3$ near the center—though higher powers are equally valid for regularising the spacetime. If experimentalists can realize these two conditions through the ratio $v_r/c_{\rm sw}$ by any means, the resulting system constitutes a valid UCO spacetime, even if it does not correspond to any previously studied analytic model. The mass function can then be extracted directly from the measured ratio, and the QNM and echo spectrum are analyzed accordingly. This *bottom-up* approach complements the *top-down* strategy adopted in this paper, where a specific $m(r,l)$ is prescribed and everything is derived from it. Both are legitimate, and it is ultimately the experimentalists who must decide which is more tractable given the constraints of the laboratory.

## ACKNOWLEDGMENTS

The authors wish to thank Leonardo Solidoro, Silke Weinfurtner, Sam Patrick, and Ralf Schützhold for illuminating discussions and suggestions. We thank an anonymous referee for their useful comments, which led to a significant improvement of our discussion of limiting factors to the experimental realization, that we summarize in Appendix H.

## DATA AVAILABILITY

There are no publicly available research data or software supporting this manuscript. Requests for further information or data should be sent to the authors.

## APPENDIX A: CONSISTENTLY TREATING *r*-DERIVATIVES AT THE FLUID SURFACE

Most authors forget that $v_r(r)$ on the surface is actually $v_r(r, z = h(r))$, which means that we cannot simply treat partial $r$-derivatives of the radial velocity restricted to the surface as total $r$-derivatives, like we also have been doing in this paper. In Ref. [35], this problem is evaded by simply assuming that the value of the radial velocity on the surface $v_r(r, z = h(r))$ equals the average over $z$ of $v_r(r,z)$ in the full volume of the fluid [see their Eq. (17)]:

$$v_r(r, z = h(r)) \approx \frac{1}{h(r)} \int_0^{h(r)} v_r(r,z) dz \qquad \text{(A1)}$$

This average is clearly no longer $z$ dependent. We could also have made this assumption of a slowly varying $v_r(r,z)$ as a function of $z$, which is quite reasonable for a fluid with negligible viscosity. In this appendix, we want to show how to treat $r$-derivatives of $v_r$ consistently and see why $\partial_r v_r(r, z = h(r)) \approx \frac{d}{dr} v_r(r, z = h(r))$ in our setting, justifying our slightly less rigorous treatment of $r$-derivatives in the main text. Briefly put, the extra terms that appear when one treats $r$-derivatives consistently, for which one must also use the irrotationality condition, are negligible because they contain $h''h'$ and $h'^2$ terms. To our knowledge, we are the first authors to make this statement explicit.

The total $r$-derivative of $v_r(r,z)$ as restricted to the surface $z = h(r)$, that we call short-hand $D$, is

$$\frac{d}{dr}[v_r(r,z)|_{z=h(r)}] = v_r' + \frac{\partial v_r}{\partial z} h' \equiv D, \qquad \text{(A2)}$$

where we continue to denote *partial* $r$-derivatives with a prime. Clearly, a partial $z$-derivative appears as well. To express everything with $r$-derivatives, we will use the continuity equation (28), the kinematic boundary condition (30), and one of the consequences of irrotationality, namely, $\frac{\partial v_r}{\partial z} = \frac{\partial v_z}{\partial r}$.

If we compute now the total $r$-derivative of $v_z(r,z)$ as restricted to the surface $z = h(r)$, we obtain

$$\frac{d}{dr}[v_z(r,z)|_{z=h(r)}] = \frac{\partial v_z}{\partial r} + \frac{\partial v_z}{\partial z} h' = \frac{\partial v_r}{\partial z} - \left(v_r' + \frac{v_r}{r}\right) h', \qquad \text{(A3)}$$

where in the second step we have inserted irrotationality and continuity. On the other hand, for the same quantity, considering the kinematic boundary condition $v_z(r,z)|_{z=h(r)} = v_r h'$, we find

$$\frac{d}{dr}[v_z(r,z)|_{z=h(r)}] = \frac{d}{dr}[v_r h'] = Dh' + v_r h''. \qquad \text{(A4)}$$

Setting these two results for $\frac{d}{dr}[v_z(r,z)|_{z=h(r)}]$ equal, we obtain the equation

$$Dh' + v_r h'' = \partial_z v_r - \left(v_r' + \frac{v_r}{r}\right) h'. \qquad \text{(A5)}$$

Along with Eq. (A2), we have two equations, whose solutions yield the two partial derivatives $v_r' = \partial_r v_r$ and $\partial_z v_r$ expressed through $D$, which is the actual derivative one can compute given $v_r$ on the surface as a certain function of $r$. For example, taking $v_r(r,z)|_{z=h(r)} = -\frac{Ur}{2h(r)}$ as in Eq. (46), we can easily compute $D = -\frac{U}{2h} + \frac{Urh'}{2h^2}$, but note that this is not $v_r'$, which is precisely the point of this appendix.

Expressing $\partial_z v_r$ through $D$ and $v_r'$ and inserting it into Eq. (A2), we finally obtain

$$v_r' = D - Dh'^2 + v_r h'' h' + v_r' h'^2 - \frac{v_r}{r} h'^2. \tag{A6}$$

In the bulk text of this paper, we argue at length why both $h'$ and $h''$ (made dimensionless) are small quantities and why we neglect terms of quadratic and higher order in those quantities. Hence, in this approximation, $v_r' \approx D$, and thus our considerations in the main text of this work are valid.

## APPENDIX B: SLOWLY ROTATING ANALOG SPACETIME

It is quite straightforward to obtain from the general static spherically symmetric metric (1) its slowly rotating equivalent in Lense-Thirring form. For this, one simply adds the standard Lense-Thirring $g_{t\phi}$ component, thus obtaining the Lense-Thirring metric. It describes the spacetime around a slowly rotating spherically symmetric object, here with an $r$-dependent mass function $m(r)$ and with angular momentum $m(r)a$ [53]:

$$ds^2_{\mathrm{GR}} = -\left(1 - \frac{2m(r)}{r}\right)dt^2 + \left(1 - \frac{2m(r)}{r}\right)^{-1} dr^2 - \frac{4m(r)a\sin^2\theta}{r} dt d\phi + d\Omega^2. \tag{B1}$$

This metric is purposely constructed linear in the small dimensionless quantity $a$. To draw comparisons to the acoustic metric, we need to put it in PG form as in Eq. (21). Here as well, the end result of the construction amounts to adding the Lense-Thirring slow-rotation term, but it is a useful exercise (left to the reader) to go through the transformation explicitly, consistently neglecting terms of $\mathcal{O}(a^2)$ and higher. The result is [54]

$$ds^2_{\mathrm{GR}} = -\left(1 - \frac{2m(r)}{r}\right)dt^2 + 2\sqrt{\frac{2m(r)}{r}} dr dt + dr^2 - \frac{4m(r)a\sin^2\theta}{r} dt d\phi + d\Omega^2. \tag{B2}$$

We now compare this to the acoustic metric, where again, to remain linear in $h'$, the terms containing $v_z$ and/or $dz$ are neglected, but now we allow for $v_\phi \neq 0$. Also, we have again performed the coordinate transformation $dt \to dt_{\mathrm{AC}} \equiv c_{\mathrm{sw}}(r)dt$, desynchronizing clocks to make the connection between the two metrics. The line element then takes the form

$$ds^2_{\mathrm{AC}} = -\left(1 - \frac{v_r^2}{c_{\mathrm{sw}}^2} - \frac{v_\phi^2}{c_{\mathrm{sw}}^2}\right) dt^2_{\mathrm{AC}} - 2\frac{v_r}{c_{\mathrm{sw}}} dr dt_{\mathrm{AC}} - 2\frac{v_\phi}{c_{\mathrm{sw}}} r d\phi dt_{\mathrm{AC}} + dr^2 + r^2 d\phi^2. \tag{B3}$$

It becomes immediate that the presence of a $g_{t\phi}$ component in both metrics induces the new correspondence between fluid and gravitational quantities,

$$\frac{v_\phi}{c_{\mathrm{sw}}} r = \frac{2m(r)a}{r}, \tag{B4}$$

where we have already restricted the quantities to the equatorial plane of the gravitational spacetime $\theta = \pi$. Consequently, at $r \to 0$, we expect the following behavior:

$$\frac{v_\phi}{c_{\mathrm{sw}}} = \frac{ar}{l^2} \propto r. \tag{B5}$$

This will be consistent with the behavior $v_\phi = J/r$ expected from an irrotational fluid only if the leading behavior at $r \to 0$ of the wave speed is $c_{\mathrm{sw}} = \sqrt{\tilde{g}(r)h(r)} \propto r^{-2}$. The behavior $\frac{v_\phi}{c_{\mathrm{sw}}} \propto r$ is in agreement, if a constant sound speed is assumed at the center, with the "solid-body rotation term" $\Omega r \subset v_\phi$ that is considered sometimes as a first-order correction to the dispersion relation [37]. However, the authors state that this term in their vortex drain setup is relevant only for large radii (it being linear in radius), whereas in our case to realize a slowly rotating spherically symmetric UCO we need it to be dominant at the center.

It seems that it is not trivial to realize a slowly rotating spacetime in a draining vortex experiment with an irrotational fluid. Whereas in the case of radial velocity we have some experimental control of the equations through the radial profile of the drainage velocity at the bottom $v_{z=0}(r)$ that enters the (cont. + k.b.) equation (43), we have much less room for tuning $v_\phi$.

To conclude, let us note that the nonvanishing azimuthal velocity does not change anything in the considerations regarding the radial velocity. The identification $\frac{v_r(r)}{c_{\mathrm{sw}}(r)} = -\sqrt{\frac{2m(r)}{r}}$ persists without issue, since the azimuthal velocity appears squared in the $g_{tt}$ term and is thus neglected, being $\mathcal{O}(a^2)$. The same is valid in the Bernoulli equation; the azimuthal velocity appears only squared in the term $\frac{1}{2}v_\phi^2$ and is thus neglected, leaving the Bernoulli equation unchanged. Because of axisymmetry, $\partial_\phi = 0$ for any quantity, and therefore the continuity equation also remains unchanged.

## APPENDIX C: CENTRIFUGAL FORCE ON THE FREE SURFACE OF A FLUID

It is well known that the speed of water waves on the surface of a water body with a constant background height $h_0$ in the shallow-water regime is $c_{\mathrm{sw}} = \sqrt{gh_0}$ with the constant gravitational acceleration $g$. The shallow-water regime is characterized by a water height/depth much smaller than the wavelength of the perturbations and is a good approximation when (i) surface tension is negligible and (ii) at the boundaries of the water body there is nothing to cause a nonzero boundary velocity $\mathbf{v}_0$ and a radial flow $v_r(r)$. This is a standard regime, for example, in geophysical situations in shallow waters, and then the Bernoulli equation simply becomes $h_B(r) = h_0$.

Here, this is not true, in general. As was explained, a nonzero boundary velocity $\mathbf{v}_0$ and $v_{z=0}(r)$ cause a fluid flow, which in turn causes a variable height profile. Thus, the fluid height is not constant also in the background, but the rate of change of the height is small, since we also want $h'_B(r) \ll 1$, so, calling $\tilde{h}(r)$ the deviation from constancy, we can write

$$h_B(r) = h_0 + \tilde{h}(r), \tag{C1}$$

Note again that, although $\tilde{h}(r) \ll h_0$, this expansion around the constant value $h_0$ at the center of the container is still an expansion for the background height; these are not the waves which “feel” the acoustic metric. Those would be a perturbation $\delta h(r)$ on top of this background,

$$h^{(1)}(r) = h_B(r) + \delta h(r), \tag{C2}$$

where the superscript “(1)” indicates that the quantity is first order in small wave perturbations. The difference between what is considered a wave perturbation and a perturbation on top of the background is somewhat a matter of convention: higher-wavelength disturbances are conventionally seen as part of the average bulk motion, whereas lower-wavelength disturbances are conventionally seen as acoustic [16]. This classification of higher-wavelength vs lower-wavelength perturbations is with respect to some conventionally chosen intermediate boundary value; one can still be in the shallow water regime for the “lower-wavelength” acoustic perturbations because their wavelength can still be much bigger than the fluid height.

The nonconstant background fluid height $h_B(r)$, that we now rename back to $h(r)$ for ease of notation, will exhibit some “curvature” owing to the nonvanishing of its derivatives. This means that fluid elements traveling along the surface will experience fictitious forces (accelerations). Among those, it is easiest to understand and identify the centrifugal acceleration, which is why we want to illustrate how it comes about in this appendix. The small centrifugal acceleration will be added on top of the gravitational acceleration pointing strictly downward; see Fig. 4. Whenever the fluid element travels along a segment of the fluid surface with concave/convex curvature, we expect the resulting effective gravitational acceleration to be smaller/larger than the constant gravitational acceleration $g$.

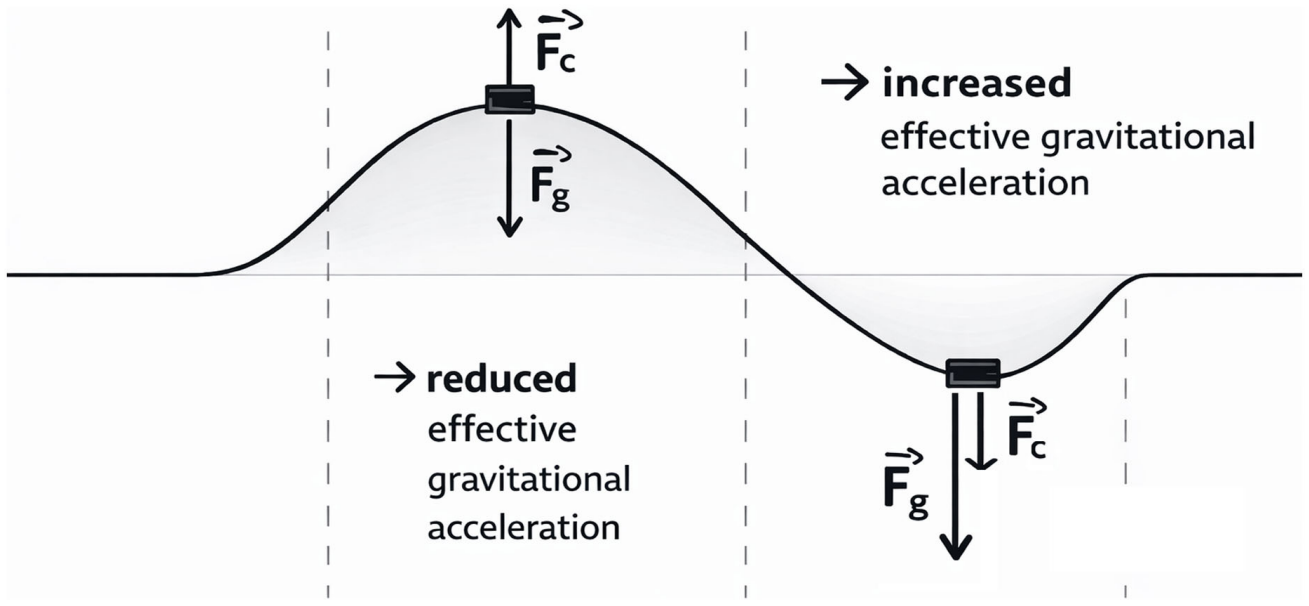

FIG. 4. Illustration of the centrifugal force felt by fluid elements in addition to the gravitational force. If absorbed into an effective gravitation acceleration $\tilde{g}$, the centrifugal acceleration leads to a reduction/increase in $\tilde{g}$ with respect to the usual constant value $g$ for concave/convex curvature of the surface of the fluid.

We recall the effective gravitational acceleration from the main text, taken from Ref. [35],

$$\tilde{g}(r) = g + \frac{D^2}{Dt^2} h(r) = g + v_r^2 h'' + v_r v_r' h', \tag{C3}$$

where $\frac{D^2}{Dt^2} = (\partial_t + \vec{v} \cdot \vec{\nabla})^2$ is the second Lagrangian derivative. Indeed, the nonconstant term corresponds to the fictitious accelerations (forces per mass) experienced by the noninertial observers traveling with the fluid elements: the linear fictitious force and the centrifugal, Coriolis, and Euler forces.

Let us consider the appearance of a centrifugal acceleration term as an example, since it is particularly simple to identify. Note that the radius of curvature $R_h(r)$, i.e., the radius of the circular arc that best fits the curve $h(r)$ at a certain point $r$, is given by [55]

$$R_h(r) = \left| \frac{(1 + h'(r)^2)^{3/2}}{h''(r)} \right| \simeq \left| \frac{1}{h''(r)} \right|, \tag{C4}$$

where in the second line we have neglected the term quadratic in $h'$ in the numerator to consistently remain linear in this small quantity. Recalling, then, from classical mechanics, that a centrifugal acceleration along a curve is given by $v^2/R$, where $v$ is the velocity along the curve and $R$ is the curvature radius of the curve, we immediately identify the first extra term $v_r^2 h''$ appearing in the effective gravitational acceleration. Indeed, for a concave height profile ($h'' < 0$), we obtain $h'' = -R_h^{-1}$, and the resulting negative centrifugal term reduces the effective gravitational acceleration, as expected. In contrast, for a convex height profile ($h'' > 0$), we obtain $h'' = +R_h^{-1}$, and the resulting positive centrifugal term increases the effective gravitational acceleration, in agreement with our simple graphic argument above.

## APPENDIX D: CONSISTENCY BETWEEN TWO SECOND-ORDER ODEs

In this appendix, we provide a simple pedagogical example of a consistency analysis for two second-order ODEs for $h(r)$. Let us take the following ODEs:

$$h''(r) = h'(r) \tag{D1a}$$

$$h''(r) = 2h'(r). \tag{D1b}$$

The first one is solved by $h_1(r) = A_1 e^r + B_1$, and the second one is solved by $h_2(r) = A_2 e^{2r} + B_2$. These solutions are in general not compatible with each other. However, imposing the consistency constraint

$$h'(r) = 2h'(r), \tag{D2}$$

which is solved by $h(r) = \text{const}$, we see that the original second-order ODEs have a common solution $h(r) = B_1 = B_2$ on a reduced common space. If we had first solved this compatibility constraint equation and then inserted the resulting constant solution into the two original equations, this would have also, equivalently, eliminated the constants $A_1$ and $A_2$, setting them to zero for consistency between the two equations.

## APPENDIX E: NONFLAT BOTTOM

In Sec. V D, we explained that in the flat-bottom surface-wave fluid system we have one variable too few to have a trivially consistent (as opposed to overdetermined) system of ODEs. This was used as an argument to consider configurations in which both $h''$ and $h'$ can be neglected. Although this reduced the ODEs to algebraic equations, it did not change the number of equations and variables: we still have three equations (two fluid and one identification equation) and two variables. Nevertheless, in Sec. VI B, we found a way to realize the central region of the UCO/RBH spacetime with the approximation of a constant $h = h_0$; what happened there is that the fluid equations were satisfied, as they must be by nature, and, given the approximation, the identification equation was fulfilled "by chance." In a sense, we did not impose it, so we only had the two fluid equations for the two fluid variables, and thus the problem of overdetermination of the equations was not really present.

To have more control and not rely on "chance," there might be another way to achieve a greater control of the system: varying the bottom profile of the fluid container. Indeed, this is equivalent to adding a third variable to the system of three equations, because it allows direct experimental control of $\tilde{g}(r)$ which will, unsurprisingly, depend on the bottom profile [39]. Following closely these authors, we now want to lay out some preliminary considerations in this regard, before delegating to future work the full exploration of the potential of the non-flat-bottom approach for our purposes. This would be a valuable contribution to the analog gravity community, as after the work of Ref. [39] theoretical considerations on varying the fluid bottom profile has been incredibly scarce if not nonexistent.

Let us assume the bottom of the fluid tank is defined by the function $Z = f(r)$, where $Z$ is a cylindrical coordinate that is related to the coordinate $z$ that we are already familiar with via the correspondence $Z = f(r) \leftrightarrow z = 0$. Then, to lowest order in the vertical height $z$ above the bottom of the fluid, its deviation from flatness is described by the metric[11]

$$d\boldsymbol{r}^2_{(0)} \equiv \eta^{(0)}_{ij} dx^i dx^j = dz^2 + (1 + f'(r)^2) dr^2 + r^2 d\phi^2. \tag{E1}$$

The sound velocity is also modified, because the curvature of the bottom also influences the effective gravitational acceleration $\tilde{g}(r)$. Isolating only the contribution of the curvature of the bottom, i.e., neglecting the extra terms due to fictitious forces at the fluid surface in Eq. (36), we obtain

$$c^2_{\text{sw}}(r) = \tilde{g}(r) h(r) = \frac{g h(r)}{\sqrt{1 + f'(r)^2}}. \tag{E2}$$

The (cont. + k.b.) equation (33) is also changed by the appearance of the factor $(1 + f'(r)^2)$, becoming [39]

$$\frac{1}{r} \left( h r v^r \sqrt{1 + f'(r)^2} \right)' = v^{z=0}(r). \tag{E3}$$

Essentially, this equation is the one we had before, but with $h(r) \to h(r)\sqrt{1 + f'(r)^2}$. This substitution will be valid also for the solutions of this equation. For example, the solution we found close to the central drain $v^r(r) = -\frac{Ur}{2h}$, with a nonflat bottom becomes $v^r(r) = -\frac{Ur}{2h\sqrt{1+f'(r)^2}}$.

Note that we have now properly written velocity components with upper indices, them being components of a contravariant vector. When the spatial metric of the bottom of the fluid was flat, this was irrelevant, so for ease of notation, we "wrongly" wrote them with lower indices. This becomes important, for example, when we compute the absolute value of the 3-velocity vector with the nonflat bottom metric

[11]This is already the third metric that we mention that is relevant in this analog gravity setup; let there be no confusion between (i) the above metric expressing in differential-geometric language the fact that the tank bottom is curved, (ii) the effective acoustic (sonic) metric that is felt by the surface-gravity waves only, (iii) the gravitational metric that is to be simulated by the above effective acoustic metric, and (iv) the "real" spacetime metric that the laboratory is immersed in, which we consider irrelevant here and is typically taken as flat.

$$\mathbf{v}^2 = \eta^{(0)}_{ij} v^i v^j = (1 + f'(r)^2)(v^r)^2 + r^2(v^\phi)^2, \tag{E4}$$

where $(v^z)^2 = (v^r h')^2$ has been neglected as in the bulk text. Note that, unlike in Sec. VA, where $v^\phi$ had the dimension of a linear velocity, length/time, here it has dimensions of an angular velocity, i.e., 1/time to stay in line with the notation of Ref. [39].

This modifies the Bernoulli equation. Setting $v^\phi = 0$ now as before yields

$$(1+f'(r)^2)(v^r(r)^2-(v^r_0)^2)+2g\frac{h(r)-h_0}{\sqrt{1+f'(r)^2}}+2gf(r)=0. \tag{E5}$$

How does the presence of $f(r)$ affect the identification between fluid and gravitational quantities? The effective $2+1$ acoustic metric, up to conformal factors, written after the desynchronization of clocks $dt \to dt_{\mathrm{AC}} = c_{\mathrm{sw}}(r)dt$ conforming to (22), is

$$ds^2_{\mathrm{AC}} = -\left(1-(1+f'^2)\frac{(v^r)^2}{c^2_{\mathrm{sw}}}\right)dt^2_{\mathrm{AC}} - 2(1+f'^2)\frac{v^r}{c_{\mathrm{sw}}}dr dt_{\mathrm{AC}} + (1+f'^2)dr^2 + r^2 d\phi^2. \tag{E6}$$

It becomes clear that, to reproduce the typical form $g_{rr} = 1$ of a gravitational metric in PG coordinates, we need to perform another coordinate transformation that experimentalists need to keep in track of: $dr \to dr_{\mathrm{AC}} = \sqrt{1+f'(r)^2}dr$, that is,

$$r(r_{\mathrm{AC}}) = r_{\mathrm{AC}}(r)^{-1} = \left(\int^r dr\sqrt{1+f'(r)^2}\right)^{-1}, \tag{E7}$$

where the inversion of the function here was denoted in a short-hand way with subscript “-1”. After this transformation, the acoustic metric becomes

$$ds^2_{\mathrm{AC}} = -\left(1-(1+f'(r(r_{\mathrm{AC}}))^2)\frac{(v^r(r(r_{\mathrm{AC}})))^2}{c_{\mathrm{sw}}(r(r_{\mathrm{AC}}))^2}\right)dt^2_{\mathrm{AC}} - 2\sqrt{1+f'(r(r_{\mathrm{AC}}))^2}\frac{v^r(r(r_{\mathrm{AC}}))}{c_{\mathrm{sw}}(r(r_{\mathrm{AC}}))}dr_{\mathrm{AC}}dt_{\mathrm{AC}} + dr^2_{\mathrm{AC}} + r(r_{\mathrm{AC}})^2 d\phi^2. \tag{E8}$$

This allows us to consistently draw the correspondence between fluid and gravitational quantities,

$$(1+f'(r(r_{\mathrm{AC}}))^2)^{3/4}\frac{v^r(r(r_{\mathrm{AC}}))}{\sqrt{gh(r(r_{\mathrm{AC}}))}} = -\sqrt{\frac{2m(r_{\mathrm{AC}})}{r_{\mathrm{AC}}}}. \tag{E9}$$

The identification is made more complicated due to the need to invert $r_{\mathrm{AC}}(r)$ to obtain $r(r_{\mathrm{AC}})$, but it is in principle possible. We have ended up with three equations: Bernoulli, (cont. + k.b.), and the identification equation and three quantities $(v^r, h, f)$. Since $f(r)$ is under experimental control, much like $v^{z=0}(r)$, it can be chosen such that the identification equation is fulfilled, given the solution (50) of the (cont. + k.b), which gives $v^r$ expressed via $\sqrt{1+f'(r)^2}$ and $h(r)$,

$$v^r(r) = \frac{1}{rh(r)\sqrt{1+f'(r)^2}}\int dr r v_{z=0}(r) + \frac{B}{rh(r)\sqrt{1+f'(r)^2}}. \tag{E10}$$

Then, one can insert the identification equation into the Bernoulli equation, eliminating $v^r$ from it, obtaining

$$-(1+f'(r(r_{\mathrm{AC}}))^2)^{3/2}\frac{(v^r_0)^2}{gh(r(r_{\mathrm{AC}}))} + \frac{2m(r(r_{\mathrm{AC}}))}{r(r_{\mathrm{AC}})} + 1 - \frac{h_0}{h(r(r_{\mathrm{AC}}))} + \sqrt{1+f'(r(r_{\mathrm{AC}}))^2}\frac{f(r(r_{\mathrm{AC}}))}{h(r(r_{\mathrm{AC}}))} = 0. \tag{E11}$$

This, along with the identification equation, with Eq. (E10) inserted to eliminate $v^r$, forms a system of two nonlinear first-order ODEs for $f(r)$ and $h(r)$. This should, in principle, ensure enough freedom in tuning the solutions via the experimentally controlled quantities $f(r)$ and $v^{z=0}(r)$ in such a way that virtually any $m(r)$ can be reproduced. Verifying this would require numerical work which is beyond the scope of this paper.

## APPENDIX F: SOLUTION TO BERNOULLI EQUATION FOR THE GLOBAL GRADIENT DRAINAGE SETUP

Here, we go through the steps necessary to demonstrate the inexistence of a positive solution in the first case of Eq. (54), which is the least trivial one. Let us denote the constant terms in the Bernoulli equation $2E = 2gh_b + v^2_{r,b}$ and call $F(r) = h^2(r)v^2_r(r) = (-\frac{Wr^{a+1}}{(a+2)b^a} + \frac{B}{r})^2 \geq 0$. Then, the Bernoulli equation becomes

$$2h^2(gh-E) = -F(r) \leq 0. \tag{F1}$$

This yields the important constraint $h \leq E/g$. However, for large $r$, $F(r) \to \frac{W^2 r^{2(a+1)}}{(a+2)^2 b^{2a}} \to \infty$ in the $a > -1$ case, forcing the positive $h(r)$ also to grow unboundedly, which is in contradiction to the upper bound we found on $h$. The “forcing” term $F(r)$ grows faster than the cubic can accommodate without violating positivity of $h(r)$, and hence there is no admissible large-$r$ solution. Physically, this can be interpreted as follows: the magnitude of the drain velocity does not fall off fast enough toward infinity, so the fluid ends up reaching the bottom of the container, i.e., $h(r) = 0$. Since it physically cannot be continued to negative values, the fluid simply stops there.

## APPENDIX G: DYNAMIC UCO SPACETIMES WITH TIME-DEPENDENT ADM MASS

In this Appendix, we want to see whether the identification (56) between ADM mass $M$ and drainage velocity $W$ at an intermediate boundary $r = b$ can be used to generalize our considerations to time-dependent spacetimes.

In the Introduction, we mentioned bouncing geometries with an inner and outer apparent horizon forming and disappearing in finite time. The former suffers from a blueshift amplification of scalar perturbations that is finite, but very large [15]. Observationally, this would manifest as a very high-energy burst that is in principle observable at infinity. This is another type of "instability" that may be testable in analog gravity.

Interestingly, such a geometry can be realized using the Hayward RBH/BHM model with a time-dependent ADM mass $M(t)$ [15]. This can also be done with a time-dependent realization of the regularization parameter $l(t)$, due to the degeneracy between $l$ and $M$ in simple UCO geometries like the Hayward model. In analog gravity, this could be achieved in a setup very similar to ours, but with a time-dependent draining velocity $U(t_{\mathrm{AC}})$ or $W(t_{\mathrm{AC}})$.

There is a caveat to this naive generalization to the nonstationary case. A time-varying draining velocity would lead to a time-dependent flow, which would require a completely different analysis that takes into account $\partial_t \phi$ term in the Bernoulli equation (15). This caveat could be circumvented by imposing a "top-hat" variation of the draining velocity, i.e., an almost instantaneous change of the parameter $l$ or $M$, which amounts to an effective turning on and off of the trapping. The two regimes (trapping on and trapping off) are both separately stationary, with the time dependence reduced to delta spikes that are not treated. It is beyond the scope of this work to check whether these considerations are fully self-consistent, but their experimental realization might be easier and more immediate than the generalization of the theoretical considerations, and thus potentially worth a try.

## APPENDIX H: ESTIMATES OF LIMITING FACTORS TO THE EXPERIMENTAL REALIZATION

### 1. Deviations from the idealized radial dependence of the flow

In the idealized setup, the flow is purely radial and the acoustic metric inherits an effective cylindrical symmetry. Deviations from this—arising, for instance, from $\phi$-dependent irregularities in the container geometry or by turbulence-induced deviations from a perfect laminar flow—break the effective spherical symmetry of the acoustic metric and induce coupling between different azimuthal sectors. This is a generic issue for any shallow water waves based experiment. By increasing the fluid viscosity (so as to dampen turbulence) or by adopting experimental refinements like a sufficiently symmetric and smooth basin. We are not aware of any previous experimental work, where these deviations from pure radial flow have been a fundamental obstruction.

### 2. Turbulence and boundary-layer effects

From Eq. (46), the radial flow velocity profile in the core region is $v_r = -Ur/(2h_0)$, and from Eq. (48), the analog regularization length is $l = 2h_0 c_{\mathrm{sw}}/U$, where $c_{\mathrm{sw}} = \sqrt{gh_0}$ is the surface-wave speed. In the paragraph just below Eq. (49), we use a water depth of $h_0 \approx 10$ cm for a drain velocity $U \approx 1$ m/s. These values lead to a background radial velocity near the core of order $|v_r| \sim 1$ m/s.

Although we have neglected viscosity in this work, we now show that taking viscous effects into account could lead to turbulence, which constitutes a primary practical constraint for the experimental realization. The relevant Reynolds number is

$$\mathrm{Re}_0 = \frac{|v_r| h_0}{\nu} \sim \frac{g^{1/2} h_0^{3/2}}{\nu} \sim \frac{1 \times 0.1}{10^{-6}} = 10^5, \qquad \text{(H1)}$$

for $h_0 = 10$ cm, where $\nu \approx 10^{-6}\,\mathrm{m}^2/\mathrm{s}$ is the kinematic viscosity of water and the subscript $_0$ indicates the central region. Note that the above result is valid for both regular black holes and black hole mimickers. For a regular black hole, the outer boundary of the core region can be taken to be the inner horizon where by construction $v_r = c_s$. For a black hole mimicker, one can still characterize it as the onset of the region where $v_r \sim c_s = (gh_0)^{1/2}$ (since we would like to be close to the extremal case even in the BHM branch); this is why we substituted $v_r = c_s$ in the above estimate. The same order-of-magnitude result can be obtained self-consistently by using Eqs. (46), (48), and (49). The numerical result for the fiducial values taken above places the flow beyond the laminar-turbulent transition for shallow-water basin flows ($\mathrm{Re} \sim 10^3$), and thus the flow would be highly turbulent.

We see that working in an even shallower regime ($h_0 \sim 1$ cm) brings $Re_0 \sim 10^3$ closer to the laminar-turbulent transition, which is still marginally problematic. The corresponding $l$ and $M$ will be centimeter sized. To solve this problem, one could use, as proposed also in Ref. [56], silicon oil as a fluid, which is 20 times more viscous than water $\nu_{oil} = 20\nu$. This leads to a decrease in $\mathrm{Re}_0$ by 1 order of magnitude, i.e., $\mathrm{Re}_0 \sim 100$ for silicon oil.

Unfortunately, an increased viscosity could lead to problems stemming from another practical constraint, namely, boundary-layer effects. Viscous boundary-layer effects at the basin floor are characterized by a layer thickness $\delta \sim \sqrt{\nu/\omega}$. For water, we obtain $\delta \sim 1$ mm for wave frequencies $\omega \sim 1$ rad/s (a typical wave frequency for wave excitations in this system [57]), which is small compared to $h_0$ and constitutes only a minor correction. However, for silicon oil, one gets $\delta \sim 5$ mm, which would be a 5% correction for a fluid depth $h_0 \approx 1$ cm.

### 3. Robustness of inner light ring position to fluctuations in fluid depth

From Eq. (48), the analog regularization length scales as $l \propto h_0^{3/2}$ at fixed drainage rate $U$, and the inner light ring radius (49) satisfies $r_{\rm ILR} \sim M \sim l$, so that $r_{\rm ILR} \propto l \propto h_0^{3/2}$. A straightforward variation calculation gives

$$\frac{\delta r_{\rm ILR}}{r_{\rm ILR}} = \frac{3}{2}\frac{\delta h_0}{h_0}. \tag{H2}$$

From Eq. (46), at fixed $r$ and fixed $U$, the radial velocity satisfies $v_r \propto h_0^{-1}$, so that $\delta v_r / v_r = -\delta h_0 / h_0$. This gives the equivalent expression

$$\frac{\delta r_{\rm ILR}}{r_{\rm ILR}} = -\frac{3}{2}\frac{\delta v_r}{v_r}. \tag{H3}$$

The sensitivity coefficient is $3/2$ in both cases, arising entirely from the $h_0^{3/2}$ scaling of $l$. This means that percent-level control of the fluid depth is sufficient to stabilize the inner light ring position to 1.5%.

### 4. Blueshift amplification before onset of dispersive corrections

The shallow-water surface-gravity wave dispersion relation is

$$\omega^2 = gk \tanh(kh_0), \tag{H4}$$

which reduces to the nondispersive limit $\omega \approx c_{\rm sw} k$ for $kh_0 \ll 1$. This is precisely the regime of validity of the shallow-water regime. Thus, dispersive corrections become significant when $kh_0 \sim 1$, i.e., at a cutoff wave number $k_{\rm disp} \sim h_0^{-1}$ and corresponding frequency [40]:

$$\omega_{\rm disp} \sim \sqrt{g/h_0}. \tag{H5}$$

For $h_0 = 1$ cm, $\omega_{\rm disp} \approx 31$ rad/s ($f \approx 5$ Hz); for $h_0 = 10$ cm, $\omega_{\rm disp} \approx 10$ rad/s ($f \approx 1.6$ Hz).

The maximum achievable blueshift factor before dispersion starts dominating is therefore

$$\frac{\omega_{\rm disp}}{\omega_0} \sim \frac{\sqrt{g/h_0}}{\omega_0}, \tag{H6}$$

where $\omega_0$ is the initial mode frequency. For $\omega_0 \sim 1$ rad/s (again, taking a typical frequency for surface waves in this system) and $h_0 = 1$ cm, this gives a blueshift factor of ~30, whereas for $h_0 = 10$ cm, the blueshift factor is ~10.

It is important to stress that, even when dispersion starts dominating, blueshift does not necessarily stop. We conjecture that it simply may not be exponential anymore (near the inner horizon, an outgoing mode undergoes exponential blueshift $\omega(t) \propto e^{|\kappa_-| t}$, where $\kappa_-$ is the inner surface gravity) but, for example, linear, as found in a different context in Ref. [17]. Because of the limitations of the surface-gravity wave approach, and since it is beyond the scope of the paper, we will not evaluate further whether the taming of the instability actually happens and in what way, in this analog platform.

### 5. Viscous dissipation timescale versus echo time delay

We want to assess whether the central analog potential well illustrated in Fig. 1 is sufficiently deep to support long-lived modes before would dissipation in the fluid starts to dominate. We do this by comparing the echo time delay $T_{\rm echo}$, which sets the timescale for long-lived modes phenomenology, with the viscous dissipation timescale $\tau_{\rm diss}$.

The echo time delay is dominated by the interior light-ray back-and-forth travel time [30]

$$T_{\rm int} = \frac{2}{c_{\rm sw}} \int_0^{r_{\rm ph}} \frac{dr}{f(r)}, \tag{H7}$$

where $f(r) = 1–2m(r)/r$ with the full Hayward mass function $m(r) = Mr^3/(r^3 + 2Ml^2)$ and $r_{\rm ph} \approx 3M$ is the outer photon sphere radius. In the mimicker branch ($l > l_*$), $f(r) > 0$ everywhere, so the integral is well defined and finite. We evaluate it numerically for representative values of $l/l_*$:

| $l/l_*$ | $\min_r f(r)$ | $T_{\rm int}[M/c_{\rm sw}]$ |
|---|---|---|
| 1.05 | 0.032 | 48.3 |
| 1.10 | 0.062 | 34.7 |
| 1.50 | 0.237 | 16.3 |
| 2.00 | 0.370 | 11.8 |

As expected, $T_{\rm int}$ diverges as $l \to l_*^+$, reflecting arbitrarily strong mode trapping near the extremal limit, indicating horizon formation.

For the viscous dissipation timescale in the shallow-water limit, we can estimate a bottom friction timescale as

$$\tau_{\rm diss} \sim \frac{h_0^2}{\nu}, \tag{H8}$$

where $\nu \approx 10^{-6}\ {\rm m^2/s}$. This gives $\tau_{\rm diss} \approx 100$ s for $h_0 = 1$ cm and $\tau_{\rm diss} \approx 10000$ s for $h_0 = 10$ cm.

Converting $T_{\rm int}$ to laboratory units for the representative case, $l = 1.1 l_*$ (so $T_{\rm int} \approx 34.7 M/c_{\rm sw}$), with $h_0 = 10$ cm leading to $l = 20$ cm, $M_{\rm analog} \approx 25$ cm, and $c_{\rm sw} \approx 1$ m/s or with $h_0 = 1$ cm leading to $l = 0.6$ cm, $M_{\rm analog} \approx 0.8$ cm, and $c_{\rm sw} \approx 0.3$ m/s,

$$T_{\mathrm{int}} \approx 34.7 \times \frac{0.25\ \mathrm{m}}{1\ \mathrm{m/s}} \approx 9\ \mathrm{s} \quad \text{or} \quad T_{\mathrm{int}} \approx 34.7 \times \frac{0.008\ \mathrm{m}}{0.3\ \mathrm{m/s}} \approx 1\ \mathrm{s}, \tag{H9}$$

yielding a ratio $\tau_{\mathrm{diss}}/T_{\mathrm{int}} \approx 1200$ or 100 for $h_0 = 10$ cm and $h_0 = 1$ cm, respectively. This is comfortably large; many echo round trips would occur before viscous losses become significant. At the same time, the central potential well should not be too deep as to make $T_{\mathrm{int}}$ too large that echoes are no more observable within a reasonable measurement timescale, which is not the case for the present estimates.

However, we note a fundamental tension: the condition for long dissipation time favors large $h_0$, while the condition for potentially avoiding turbulence favors small $h_0$ values. This competing constraint is another inherent limitation of the shallow-water platform.